\documentclass[12pt]{article}
\usepackage{axodraw}
\usepackage{epsfig}   % fr Grafiken einbinden
\usepackage{amssymb}
\usepackage{latexsym}
\usepackage{amsmath}

\usepackage{graphics,subfigure} 

\usepackage{color}   % to use the color caption in the histograms

\newcommand{\eps} {\epsilon}

\newcommand{\tanb} {\tan\beta}
\newcommand{\mhpm} {m_{H^{\pm}}}
\newcommand{\mhone} {m_{h_{1}}}
\newcommand{\hone} {h_{1}}
\newcommand{\htwo} {h_{2}}
\newcommand{\mhtwo} {m_{h_{2}}}

\newcommand{\ra} {\rightarrow}

\def\met        {E\!\!\!\!/_T}

\newcommand{\lsim}{\raisebox{-0.13cm}{~\shortstack{$<$ \\[-0.07cm] $\sim$}}~}

\newcommand{\bea}{\begin{eqnarray}}
\newcommand{\eea}{\end{eqnarray}}
\newcommand{\beq} {\begin{equation}}
\newcommand{\eeq} {\end{equation}}

\begin{document}
\pagestyle{empty}
\begin{flushright}
IFIC/10-39 \\
October 2010 \\
\end{flushright}
\begin{center}
{\large\sc {\bf CP-violating Supersymmetric Higgs at the Tevatron and LHC}}

\vspace{1cm}
{\sc Siba Prasad Das$^{1}$ and Manuel Drees$^{2,3}$ }

\vspace*{5mm}
{}$^1${\it 
AHEP Group, Institut de F\'{\i}sica Corpuscular --
  C.S.I.C./Universitat de Val{\`e}ncia \\
  Edificio Institutos de Paterna, Apt 22085, E--46071 Valencia, Spain
}\\
{}$^2${\it Bethe Center for Theoretical Physics and Physikalisches
  Institut, Universit\"at Bonn, Nussallee 12, D--53115 
  Bonn,  Germany} \\
{}$^3${\it School of Physics, KIAS, Seoul 130--722, Korea}

\end{center}

%\maketitle

\vspace*{1cm}
\begin{abstract}

  We analyze the prospect for observing the intermediate neutral Higgs boson
  ($h_2$) in its decay to two lighter Higgs bosons ($h_1$) at the presently
  operating hadron colliders in the framework of the CP violating MSSM using
  the PYTHIA event generator. We consider the lepton+ 4-jets+ $\met$ channel
  from associate $W h_2$ production, with $W h_2 \ra W h_1 h_1 \ra \ell
  \nu_\ell b \bar b b\bar b$. We require two, three or four tagged
  $b$-jets. We explicitly consider all relevant Standard Model backgrounds,
  treating $c$-jets separately from light flavor and gluon jets and allowing
  for mistagging. We find that it is very hard to observe this signature at
  the Tevatron, even with 20 fb$^{-1}$ of data, in the LEP--allowed region of
  parameter space due to the small signal efficiency, even though the
  background is manageable. At the LHC, a priori huge SM backgrounds can be
  suppressed by applying judiciously chosen kinematical selections. After all
  cuts, we are left with a signal cross section of around 0.5 fb, and a signal
  to background ratio between 1.2 and 2.9. According to our analysis this
  Higgs signal should be viable at the LHC in the vicinity of present LEP
  exclusion once 20 to 50 fb$^{-1}$ of data have been accumulated at
  $\sqrt{s}=14$ TeV.

\end{abstract}

\newpage

\setcounter{page}{1}
\pagestyle{plain}

%%%%%%%%%%%%%%%%%%%%%%%%%%%%%%%%%%%%%%%%%%%%%%%%%%%%%%%%%%%%%%%%%%%%%%%%%%%%%%%

\section{Introduction} 
\label{intro}

The Minimal Supersymmetric Standard Model (MSSM) \cite{MSSM} requires
two Higgs doublets, leading to a total of five physical Higgs bosons.
At the tree level, these can be classified as two neutral CP--even
bosons ($\phi_1$ and $\phi_2$), one neutral CP--odd boson ($a$) and
two charged bosons. In the presence of CP violation, the three neutral
Higgs bosons can mix radiatively \cite{CPmixing0,Lee:2003nt}. The mass
eigenstates $h_1$, $h_2$ and $h_3$ with $m_{h_1} < m_{h_2} < m_{h_3}$
can then be obtained from the interaction eigenstates $\phi_1$, $\phi_2$
and $a$ with the help of the orthogonal matrix $O_{\alpha i}$,
$(\phi_1,\phi_2,a)^T_\alpha= {O_{\alpha i}}(h_1,h_2,h_3)^T_i \, ,$
which diagonalizes the Higgs boson mass matrix. $O$ depends on various
parameters of the SUSY Lagrangian.

Due to this mixing, the Higgs mass eigenstates are no longer CP
eigenstates.  Moreover, the masses of the Higgs bosons, their
couplings to SM and MSSM particles, and their decays are significantly
modified \cite{Lee:2003nt}. For example, the Higgs boson couplings to
pairs of gauge bosons are scaled by $g_{h_iVV}$ relative to the
SM. These couplings can be expressed as $g_{h_iVV} = \cos\beta\,
O_{\phi_1 i}\: +\: \sin\beta\, O_{\phi_2 i} \, ,$ where $\tan\beta$ is
the ratio of Higgs vacuum expectation values (VEVs). The magnitude of
$g_{h_2 W W}$ is directly related to the production process studied in
this paper.

In the absence of mixing between neutral CP--even and CP--odd states
the LEP experiments were able to derive absolute lower bounds of about
90 GeV on the masses of both the lighter CP--even Higgs and the
CP--odd boson \cite{adlo}.  However, in the presence of CP violation,
the LEP experiment were not able to exclude certain scenarios with
very light $h_1$. In this ``LEP hole'' $h_1$ is dominantly a CP--odd
state with almost vanishing coupling to the $Z$ boson. One then has to
search for $Z h_2$ or $h_1 h_2$ production. In part of the LEP hole,
these cross sections are suppressed by the rather large $h_2$
mass. Moreover, $h_2 \ra h_1 h_1$ decays lead to quite complicated
final states, which often yield low efficiencies after cuts.  One
LEP--allowed region has $m_{h_1}\lsim 10$ GeV, so that $\hone \ra
\tau^+ \tau^-$ is dominant; in the other, $m_{h_1} \sim 30 - 50$ GeV
so that $\hone \ra b \bar b$ is dominant. $m_{h_2}$ lies between
slightly below 90 and slightly above 130 GeV. Scenarios with even
lighter $h_2$ are excluded by decay--independent searches for $Z h_2$
production \cite{adlo,Abbiendi:2002qp,Abdallah:2004wy}. If $m_{h_2}$
is much above 130 GeV, the CP--odd component of $h_1$ becomes
subdominant, so that the cross section for $Z h_1$ production becomes
too large. Finally, the LEP hole occurs for $\tan\beta$ in between 3
and 10 \cite{adlo,Bechtle:2008jh}.

In this paper we analyze the prospect for observing a signal for the
production of neutral Higgs bosons in the second of these LEP allowed
regions. Since the $h_1 W W$ coupling is suppressed along with the
$h_1 Z Z$ coupling, we focus on $W h_2$ production, with $h_2 \ra h_1
h_1 \ra b \bar b b \bar b$ and $W \ra \ell \nu$, where $\ell$ is an
electron or muon. This process has recently been studied in
refs.\cite{Koreans,Han}, using parton--level analyses, with quite promising
results.  We instead performed a full hadron--level analysis,
including initial and final state showering as well as the underlying
event.  We will see that these effects significantly reduce the basic
kinematical efficiencies of the signal, while allowing the background
to populate new regions of phase space.  Moreover, we have expanded
the list of background processes; some of the backgrounds not
considered in \cite{Han} turn out to be sizable.  Altogether this
leads to a reduced significance to isolate the signal from the SM
backgrounds. 

The Higgs boson masses, their coupling to gauge bosons and their
branching ratios are also modified in some other models. Examples are
scenarios with spontaneous CP violation \cite{Ham:2009gu} and models
with additional Higgs singlet \cite{Chang:2005ht}. The simplest of
these is the next--to--minimal supersymmetric standard model (NMSSM)
\cite{nmssm}, the CP violating version of which has also recently been
discussed \cite{Ham2}. In all these scenarios the process we are
considering is possible, and might be useful as a discovery
channel. For example, in the CP--conserving NMSSM, the role of $h_2$
could be played by either the lightest or the second lightest CP--even
scalar. In the former case, the role of $h_1$ would be played by the
lighter CP--odd scalar, whereas in the latter case, the role of $h_1$
could also be played by the lightest CP--even scalar
\cite{Ellwanger2}.

In order to be as model--independent as possible, we chose several benchmark
points where  we simply fix the masses $m_{h_1}$ and $m_{h_2}$ as well as the
relevant product of couplings and branching fractions. In addition we
investigate a couple of benchmark points within the so--called CPX scenario of
the MSSM~\cite{Carena:2000ks}, where the masses and couplings of the Higgs
bosons can be computed in terms of the fundamental input parameters. We find
that the signal can be detectable at the LHC once 20 to 50 fb$^{-1}$ of data
have been accumulated at $\sqrt{s} = 14$ TeV; since in any case a large
integrated luminosity is needed, we do not consider scenarios with smaller
$\sqrt{s}$. The situation at the Tevatron seems hopeless due to the very small
signal.

The remainder of this paper is organized as follows. In Sec.~2 we
introduce our benchmark points, and the the numerical procedures to
estimate signal and backgrounds. In Sec.~3 results for the Tevatron
will be discussed. In particular, we mention the detector parameters,
and introduce different (combinations of) kinematical cuts, in order
to get rid of the a priori huge SM background and retain as many
signal events as possible. Sec.~4 presents a similar analysis for the
LHC. Finally we will summarize in Sec. 5.

%%%%%%%%%%%%%%%%%%%%%%%%%%%%%%%%%%%%%%%%%%%%%%%%%%%%%%%%%%%%%%%%%%%%%%%%%%

\section{Numerical analysis}
\label{numerical}

In this section we first describe our benchmark scenarios. We then
give a list of the background processes we included in our analysis,
and discuss the event generation.

\subsection{Benchmark scenarios for the signal}

In our analysis we took five different benchmark points, denoted by S1
through S5, with different intermediate Higgs boson masses ($\mhtwo$)
in between 90 and 130 GeV, while the lighter Higgs boson mass
($\mhone$) is fixed at 30 GeV; this resembles the model--independent
approach taken in ref.\cite{Han}. Note that a Higgs boson with
SM--like coupling to the $Z$ (which is equivalent to demanding an
SM--like coupling to the $W$) is excluded if its mass is below 82 GeV,
independent of its decay mode
\cite{Abbiendi:2002qp,Abdallah:2004wy}. On the other hand, in the
CP--violating MSSM values $m_{h_2}$ significantly above 130 GeV are
incompatible with a strongly reduced $Z Z h_1$ coupling \cite{adlo}.
Moreover, for $m_{h_2} > 130$ GeV, the standard decays into
$WW^*,ZZ^*$ are expected to become dominant, reducing the branching
ratio for the $h_2 \ra h_1 h_1$ decays we are interested in.

In addition we considered two benchmark points in the  CPX--scenario 
of the MSSM~\cite{Carena:2000ks}, which is defined by the following set of 
input parameters:
\begin{eqnarray}
\widetilde{M}_Q &=& \widetilde{M}_t =
\widetilde{M}_b = 500 \ {\rm GeV},\qquad
\mu  = 4 \widetilde{M}_Q \,,\nonumber\\
|A_t| &=& |A_b| = 2 \widetilde{M}_Q,\qquad
{\rm arg}(A_t) = {\rm arg}(A_b)  =  90^\circ\,, \nonumber\\
|m_{\tilde{g}}| &=& 1~{\rm TeV}\,,\qquad
{\rm arg}(m_{\tilde{g}})\ =\ 90^\circ\, .
\label{eq:CPX}
\end{eqnarray}
Here $\widetilde M_Q$ is the mass of third generation $SU(2)$ doublet
squarks, and $\widetilde M_t, \, \widetilde M_b$ are the masses of the
corresponding singlets. $\mu$ is the supersymmetric higgsino mass
parameter, $A_{b,t}$ are trilinear soft breaking parameters in the
sbottom and stop sectors, respectively, and $m_{\tilde g}$ is the
gluino mass. We work in the convention where $\mu$ is real; in the CPX
scenario, $A_{b,t}$ and $m_{\tilde g}$ are purely imaginary. The
remaining two input parameters are the charged Higgs boson mass
$\mhpm$ and $\tanb$. We take $M_{H^+} = 131.8$ GeV, and $\tanb = 4.02
\ (4.39)$ in benchmark point CPX--1 (CPX--2). We calculated the
spectrum and the couplings for these two benchmark points using {\tt
  CPsuperH}~\cite{Lee:2003nt}. Other packages like FeynHiggs
~\cite{FeynHiggs} can also be used for this purpose.  In CPX--1
$\mhone$ and $\mhtwo$ are 36 and 101.6 GeV, while in CPX--2, which is
close to the upper edge of the LEP allowed region, these masses are 45
and 102.6 GeV respectively.

\begin{table}[ht]
\begin{center}
\begin{tabular}{|c||r|c||c|c|}
\hline
& \multicolumn{2}{|c||}{mass [GeV]} &\multicolumn{2}{|c|}{$\sigma$ [pb] }\\
Scenario& $h_2$ & $h_1$ & Tevatron & LHC\\
\hline 
%%%%%%%%%%%%%%%%%%%%%%%%%%%%%%%%%
S1 & 130 & 30 & 0.090 & 1.091 \\
S2 & 120 & 30 & 0.122 & 1.402 \\
S3 & 110 & 30 & 0.163 & 1.851 \\
S4 & 100 & 30 & 0.223 & 2.472 \\
S5 &  90 & 30 & 0.315 & 3.317 \\
CPX--1 & 102 & 36 & 0.212 & 2.367\\
CPX--2 & 103 & 45 & 0.206 & 2.284 \\
%%%%%%%%%%%%%%%%%%%%%%%%%%%%%%%%%
\hline
\end{tabular}
\caption{The seven benchmark scenarios we consider. Scenarios S1
  through S5 are defined purely phenomenologically, in terms of the
  masses of the relevant Higgs bosons, whereas scenarios CPX--1 and
  CPX--2 have been obtained within the CPX set of the MSSM. The last
  two columns give the total cross sections for $W h_2$ production at
  the Tevatron ($p \bar p$ collisions at $\sqrt{s} = 1.96$ TeV) and
  the LHC ($pp$ collisions at $\sqrt{s} = 14$ TeV), assuming
  SM--strength for the $h_2 W W$ coupling.}
\label{tab:bench}
\end{center}
\end{table}

The seven benchmark scenarios are summarized in Table~\ref{tab:bench}.  All of
them satisfy $\mhtwo > 2 \mhone$ and $\mhone > 2 m_b$, so that the
$\ell$+4$j$+$\met$ signal topology that we are interested in can arise from
the associated $W\htwo$ production at hadron colliders. Note finally that a
scenario very similar to S4 can be realized in the CPX framework of the MSSM,
with $m_{H^+} = 127.9$ GeV and $\tanb = 4.31$.

The last two columns in Table~\ref{tab:bench} give the total cross
sections for $W h_2$ production at the Tevatron ($p \bar p$ collisions
at $\sqrt{s} = 1.96$ TeV) and the LHC ($pp$ collisions at $\sqrt{s} =
14$ TeV), assuming SM--strength for the $h_2 W W$ coupling. We set
the factorization scale to $Q= \sqrt{\hat s}$ (the partonic
center--of--mass energy) and used CTEQ5L \cite{cteq5} for the parton
distribution functions (PDF). The production cross section is
independent of $\mhone$. We see that going from the Tevatron to the
LHC increases the cross section by about a factor of eleven.

If the Higgs sector only contains $SU(2)$ doublets and singlets, the $h_2 W W$
coupling can only be reduced from its SM value by mixing. Moreover, for our
signal we require the $W$ boson to decay leptonically, and $h_2$ to decay into
four $b$ (anti--)quarks via two on--shell $h_1$ bosons, thus leading to $\ell
jjjj \met$ events, where $\ell = e$ or $\mu$. The cross section for this
signal topology can be expressed as
\beq  \label{effcross}
\sigma_{\rm signal}^{\rm tot} = \sigma_{SM}(p \bar p/pp \rightarrow W
\htwo) \times  C_{\htwo WW} \times 2{Br(W \rightarrow e \nu_{e})}\, 
\eeq
where we have introduced the quantity
\beq  \label{effhww}
C_{\htwo WW} =  g^2_{h_2 W W}  \times Br(\htwo \ra \hone\hone) \times
Br(\hone \ra b \bar b)^2\, .
\eeq
Here $g_{h_2 WW}$ is the $h_2 W W$ coupling in units of the
corresponding SM value, and the factor 2 is for $\rm \ell = e$ and
$\mu$.  We have taken $Br(W \rightarrow e \nu_{e}) = 0.106$. Following
ref.\cite{Han}, we set $C_{\htwo WW} = 0.50$ in our numerical
analysis. This is slightly larger than the values computed for
scenarios CPX--1 and CPX--2, which are 0.465 and 0.435,
respectively. Our results can trivially be scaled downward by the
corresponding factors; however, this correction is smaller than the
theoretical uncertainty of our leading order calculation.

\subsection{Background processes}

We just saw that our signal consists of four jets, one charged lepton,
and missing $E_T$, where the latter two components come from the decay
of an on--shell $W^\pm$ boson. In our background estimate we only
include processes that lead to the same topology, i.e. we insist on
the presence of at least one leptonically decaying real $W^\pm$ boson
in the event. We thus ignore background contributions where a jet is
misidentified as a lepton, as well as backgrounds where the entire
missing $E_T$ is due to mismeasurements or incomplete coverage of the
detector. We trust that these instrumental backgrounds are small after
cuts.

However, we do {\em not} insist on having four $b$ (anti--)quarks in
the event. In fact, we will see shortly that requiring all four $b$
jets in the signal to be tagged as such leads to a very low
efficiency, i.e. low signal rate. Moreover, there is a finite
probability that jets are mistagged, i.e. are tagged as $b-$jets even
though they do not contain a $b-$flavored hadron. We thus include {\em
  all} leading order processes that produce final states containing at
least one (leptonically decaying) $W^\pm$ boson and at least four
jets. This includes $t \bar t$ production, as well as processes where
the $W^\pm$ boson is produced directly rather than from the decay of
a top quark.

This latter class of reactions includes a large number of final states. The
largest cross sections are for final states not containing any heavy
quarks. However, these will be suppressed heavily by $b-$tagging
requirements. Processes containing at least a couple of heavy quarks in the
final state have smaller cross sections, but much higher efficiencies. In
order to obtain a reliable background estimate without having to generate huge
numbers of events, we separate the direct $W+4j$ production processes into
many categories, according to the number of heavy quarks, light quarks and
gluons in the final state. For reasons that will become clear shortly, we also
include $t \bar t b \bar b$ and $t \bar t c \bar c$ in our background
estimate. In the signal, and in all $W+4j$ backgrounds, we force the $W$ boson
to decay leptonically. Similarly, in all backgrounds containing a $t \bar t$
pair, we force one $W$ boson to decay leptonically, and the second $W$ boson
to decay into a tau--lepton or hadronically.

A complete list of the backgrounds we consider is given in
Table~\ref{tab:procs}. Processes p2 through p8 are all mixed QCD--electroweak
$W+4j$ processes, but differ in the number of $b-$quarks. Process p8, which
does not have any $b-$quarks in the final state, has by far the largest cross
section. It can be broken up into processes p8.1 through p8.9, which have
different numbers of gluons and charm (anti--)quarks in the final state; the
latter are treated separately, since charm jets have a much higher probability
to be mistagged as $b-$jets than jets originating from light flavors or
gluons. Process p8.2, which has the largest cross sections of all, is further
broken up according to the charge of the produced $W$ boson.

%%%%%%%%%%%%%%%%%%%%%%%%%%%%%%%%%%%%%%%%%%%%%%%%%%%%%%%%%%%%%%%%%%%%%%%%%% 
% List of processes
\begin{table}[h!]
\begin{center}
\begin{tabular}{|c||c|c||c|c|}
\hline
& & &\multicolumn{2}{|c|}{$\sigma$ [pb]}\\
label& final state & $j$ & Tevatron& LHC\\
\hline 
%%%%%%%%%%%%%%%%%%%%%%%%%%%%%%%%%%%%%%%%%%%%%%%%%%%%%%%%%%%%%%%
p1 & $t \bar t$ &   & 5.00 & 500.0 \\
p2 & $b \bar b b \bar b W^{\pm}$ &   & 0.015 & 0.156 \\
p3 & $b \bar b b j W^{\pm}$ & $udscg$ & $5\cdot 10^{-5}$ & 0.011 \\ 
p4 & $b \bar b c j W^{-}$ & $udsg$ & 0.152 & 33.8 \\ 
p5 & $b \bar b c \bar c W^{\pm}$ &  & 0.051 & 0.521 \\
p6 & $b \bar b jj W^{\pm}$ & $udsg$ & 5.99 & 248 \\ 
p7 & $b jjj W^{\pm}$ & $udscg$ & 0.017 & 3.32 \\
p8 & $jjjj W^{\pm}$ & $udscg$ & 447.9 & $2.93 \cdot 10^4$ \\
\hline 
p8.1 & $gggg W^{\pm}$ &  & 93.4 & 918.6 \\
p8.2 & $gggj W^{\pm}$ & $udsc$ & 206.4 & $1.97 \cdot 10^4$\\
\hline \hline 
p8.2.1 & $gggj W^{+}$ & $ds$ & 67.0 & 9302  \\
p8.2.2 & $ggg \bar j W^{+}$ & $uc$ & 31.9 & 1853 \\
p8.2.3 & $gggj W^{-}$ & $uc$ & 31.8 & 5013 \\
p8.2.4 & $ggg \bar j W^{-}$ & $ds$ & 67.2 & 1562 \\ 
\hline
$\sum_{i=1}^4$ p8.2.$i$ & & & 197.9 & $1.77 \cdot 10^4$\\ 
\hline \hline
p8.3 & $ggjj W^{\pm}$ & $udsc$ & 122.6 & 6476 \\
%\hline \hline 
%p8.3.1 & $qq \ra ggjj W^{\pm}$ & $udsc$ & 110.6 & 3150 \\
%p8.3.2 & $gg \ra ggjj W^{\pm}$ & $udsc$ & 12.0 & 3325 \\
%\hline 
%$\sum_{i=1}^2$ p8.3.i & &  & 122.6 & 6476 \\
%\hline \hline
p8.4 & $gjjj W^{\pm}$ & $udsc$ & 25.4 & 2263 \\
p8.5 & $\bar c jjj W^{+}$ & $uds$ & 0.435 & 50.9 \\
p8.6 & $c \bar c jj W^{\pm}$ & $uds$ & 0.991 & 33.3 \\
p8.7 & $c \bar c c j W^{-}$ & $uds$ & 0.094 & 15.0 \\
p8.8 & $c \bar c c \bar c W^{\pm}$ & & 0.049 & 0.471 \\
p8.9 & $jjjj W^{\pm}$ & $uds$ & 2.67 & 99.4 \\
\hline 
$\sum_{i=1}^9$ p8.$i$ & & & 443.6 & $2.76 \cdot 10^4$\\
\hline \hline
p9 & $t \bar t b \bar b$ &  & 0.0090 & 2.99 \\
p10 & $t \bar t c \bar c$ &  & 0.016 & 4.86 \\
\hline 
$\sum_{i=1}^{10}$ p$i$ & &&  454.9 & $2.84 \cdot 10^4$ \\
%%%%%%%%%%%%%%%%%%%%%%%%%%%%%%%%%%%%%%%%%%%%%%%%%%%%%%%%%%%%%%%
\hline 
\end{tabular}
\caption{List of background processes and their total cross sections
  after the pre--selection cuts (\ref{presel}). Charge conjugate final
states are included, if they differ from the listed final states, {\em
except} for processes p.8.2.$i$, $i=1 \dots 4$. The symbol $j$ stands
for different partons, as listed in the third column, where the
corresponding antiquarks are always included. See the text for further
explanations.}
\label{tab:procs}
\end{center}
\end{table} 

We used {\tt MadGraph/MadEvent v4.4.15} \cite{Maltoni:2002qb} for generating
parton level SM $W+4j$ background events and for the calculation of the
corresponding cross sections. We simulated $t \bar t$ event sample using {\tt
  PYTHIA}, assuming $m_t = 172.6$ GeV as pole mass. We again employ CTEQ5L
\cite{cteq5} parton distribution functions, with factorization and
renormalization scale given by $\sqrt{\hat s}$. We only include $u,d,s$ quarks
and gluons in the initial state, since we generate all heavy $c$ and $b$
quarks explicitly; at least in case of $b-$quarks the required transverse
momentum is only a few times larger than the mass of the quark, making the use
of $b-$quark distributions in the proton questionable. Flavor mixing has been
included where appropriate, using current values for the charged current
couplings \cite{pdg}.

The resulting cross sections are listed in Table~\ref{tab:procs}. They
have been calculated with the following kinematical cuts, also used
for the generation of events:
\bea \label{presel}
p_T^{j,b} &\geq& 5 \ {\rm GeV}\,; \nonumber \\
\eta^{j,b} &\leq& 5.0\,;  \\
\Delta R (jj,bb,bj) &\geq& 0.3\,. \nonumber
\eea
Here $\Delta R = \sqrt{(\Delta \eta)^2 + (\Delta \phi)^2}$, where
$\eta$ and $\phi$ are the pseudorapidity and azimuthal angle,
respectively. Note that some cuts on the partonic transverse momenta,
and on the separation between partons, are necessary in order to
obtain finite cross sections. We chose pre--selection cuts on the
generated partons that are much weaker than our final analysis cuts,
since showering can change the transverse momenta and separations
significantly. 

Note that the sum of subprocess cross sections in
Table~\ref{tab:procs} often does not exactly agree with the total
cross section listed first. The reason is that these numbers result
from different runs of {\tt MadGraph/MadEvent}, with the given more or
less inclusive final state. Since the total event statistics for the
sum of subprocesses is higher, and the cross section calculation
becomes more reliable when fewer different subprocesses need to be
added, we consider the summed subprocess cross sections more reliable
estimates, and use these in the estimate of the total background cross
section after pre--selection cuts, which is given in the last line of
Table~\ref{tab:procs}.

We see that going from the Tevatron to the LHC increases the raw
background cross section for process p2, which most directly resembles
our signal, by about the same factor as the signal cross
section. However, the cross sections for other background processes
increase much more rapidly. This is true in particular for the total
$t \bar t$ production cross section, process p1, which increases by
two orders of magnitude; the cross sections for processes p4 and p6,
which increase by factors of 220 and 40, respectively; and for the $t
\bar t Q \bar Q$ cross sections ($Q=b,c$), processes p9 and p10, which
increase by a factor of about 300. The smallest ratio obtains for
cross sections dominated by $u \bar d$ or $\bar u d$ annihilation,
which includes the signal as well as background processes p2, p5, p8.1
and p8.8. Here the increase is limited by the fact that the LHC is a
$pp$ collider, whereas the Tevatron is a $p \bar p$ collider.  Process
p6 receives sizable contributions from $gq$ scattering as well as $gg$
fusion; the corresponding parton fluxes increase faster than the $u
\bar d$ flux does. Process p4 receives sizable contributions from the $gs$
initial state; contributions with $gd$ initial state are suppressed by
a factor $\left| V_{cd} \right|^2  \simeq 0.053$. Finally, the
increase is largest for processes p9 and p10, which at the LHC receive
dominant contributions from the $gg$ initial state, but suffer from
small parton flux factors at the Tevatron due to the large required
partonic center--of--mass energy.

\subsection{Monte Carlo simulation}

As noted above, all parton--level events have been generated by {\tt
  MadGraph/MadEvent}. They are then passed on to {\tt PYTHIA} v.6.408
\cite{pythia}, which handles initial and final state showering and
hadronization; {\tt PYTHIA} also adds an ``underlying event'' due to
the spectator partons and their interactions. We utilize the ``old''
shower algorithm based on virtuality ordering. We use the default,
large squared shower scale $4 \hat{s}$ for the signal as well as for
inclusive $t \bar t$ production. All other background processes
typically include relatively soft particles already at the hadron
level; we therefore use a much smaller squared shower scale $\sim
0.002 \hat{s}$ for all other background processes. Initial state
radiation adds extra energy, and perhaps additional jets, to the
event. It can also give it a transverse kick, allowing partons with
relatively low transverse momentum in the partonic center--of--mass
system to produce jets that pass our acceptance cuts (which are
specified below). For our purposes final state radiation is also very
important, since it smears out invariant mass distributions, thereby
e.g. allowing the reconstructed jets from top decay to have a total
invariant mass well below the top mass. We will see below that parton
showering, which is not included in earlier parton--level analyses
\cite{Koreans,Han}, re--establishes $t \bar t$ production as most
important background after cuts. Note that showering, and hence the
choice of showering scale, is much less important for the other
background processes.

In roughly 1 to 4\% (5 to 15\%) of the events\footnote{The exact
  fraction depends on the process under consideration.}, showering
will produce at least one additional $b \bar b$ ($c \bar c$) pair. In
almost all background processes including such events would lead to
double counting, since the production of final states with multiple
heavy quarks has been treated explicitly in our list of backgrounds.
For example, producing another $b \bar b$ pair when showering an event
of type p1 would lead to an event of the type p9; producing a $c \bar
c$ pair when showering an event of type p6 would yield an event of
type p5; and so on. Vetoing these potentially double counted events is
rather important, since they can have a much larger efficiency for
passing the multi $b-$tag requirement we impose to reduce backgrounds
than events without additional heavy quarks from
showering.\footnote{In principle we could allow showering to produce
  an additional $b \bar b$ pair in processes p2, p3 and p9, since we
  do not explicitly include final states with more than 4 $b-$quarks
  in our calculation. However, since these processes already have at
  least three $b-$quarks in the final state, the small fraction of
  events that have another $b \bar b$ pair due to showering do {\em
    not} have a much higher efficiency. We thus only make a very small
  mistake, well below the accuracy of our leading order calculation,
  by throwing these events away. Analogous remarks apply to processes
  p2, p3, p4, p5, p9 and p10 regarding additional $c \bar c$
  production due to showering.} Since such events are rare, they
would also lead to very slow convergence of the MC simulation.

Finally, our simulation also includes experimental resolution smearing
for the jet angles and energies, using the toy calorimeter {\tt PYCELL}
provided by {\tt PYTHIA}. This is of some importance, since invariant
mass distributions will be used to isolate the signal. The assumed
detector characteristics differ for the Tevatron and LHC, as detailed below.

\section{Tevatron}
\label{tev_analysis}

We simulate our signal and backgrounds at Tevatron Run-II with $\sqrt s =1.96$
TeV. We base our {\tt PYCELL} model on the CDF detector \cite{Lukens:2003aq}, the
calorimeter of which covers $\rm |\eta| < 3.64$; its segmentation is $\Delta
\eta \times \Delta \phi$=$0.16 \times 0.098$. We use the same Gaussian energy
smearing for jets and leptons, with resolution
\bea \label{smearing}
{\Delta E^{j,\ell} \over E^{j,\ell}} = {75\% \over \sqrt{E^{j,\ell}}}
\oplus 5\% \quad ,  
\eea
where $\oplus$ means addition in quadrature; note that smearing of the
lepton energy is not important for our analysis. 

We reconstructed the missing energy ($\met$) from all observed
particles.\footnote{We have also calculated $\met$ from the energy
  deposition in the calorimeter cells and found consistency between
  these two methods.} We have not included any cracks in the detector
coverage in our simulations.

Jets are reconstructed using a cone algorithm, with cone size $R(j)
= \sqrt{\Delta\eta^{2}+\Delta\phi^{2}} = 0.4$.  All calorimeter cells
with $\rm E_{T,min}^{cell} \ge 1.0$ GeV are considered to be potential
candidates for jet initiator. All cells with $\rm E_{T,min}^{cell} \ge
0.1$ GeV are treated as parts of the would--be jet. Finally, jets are
required to have $\rm E_{T,min}^{j} \ge 10$ GeV, and the jets are
ordered in $E_{T}$.

Leptons ($\ell = e, ~\mu$) are selected with $E_T^{\ell} \ge 15$ GeV
and $\left| \eta^{\ell} \right| \le 2.0$. Note that our jet algorithm
also includes leptons as parts of jets. If we find a jet near a
lepton, with $\Delta R (j-\ell) \le 0.4$ and $ 0.8 \le \rm
E_{T}^{j}/E_{T}^{\ell} \le 1.2$, i.e. if the jet $E_T$ is nearly
identical to that of this lepton, the jet is removed from the list of
jets and treated as a lepton. However, if we find a jet within $\Delta
R (j-\ell) \le 0.4$ of a lepton, whose $E_T$ differs significantly from
that of the lepton, the lepton is removed from the list of
leptons. This isolation criterion should remove (most) leptons from
$b$ or $c$ decays.

The tagging of $b-$jets plays a crucial role in our analysis
\cite{Hanagaki:2005fz}. A jet with $\left|\eta^j \right| \le 1.2$ and $E_{T}^j  
\geq 15$ GeV ``matched'' with a $b-$flavored hadron ($B-$hadron), i.e.  with
$\Delta R(j,B-{\rm hadron}) < 0.2$, is considered to be ``taggable''. We
assume that such jets are actually tagged with probability $\eps_b =
0.50$. We find that our tagging algorithm agrees well with the $t \bar t$
analysis of CDF \cite{cdfttbar}.

We also modeled mistagging of non$-b$ jets as $b-$jets, treating
$c-$jets differently from those due to gluons or light quarks. A jet
with $\left| \eta^j \right| \le 1.2$ and $E_{T}^j \geq 15$ GeV matched
with a $c-$flavored hadron ($C-$hadron, e.g., a $D-$meson or
$\Lambda_c-$baryon), i.e., with $\Delta R(j,C-{\rm hadron}) < 0.2$, is
again considered to be taggable, with (mis)tagging probability
$\eps_c = 0.10$. Jets that are associated with a $\tau-$lepton, with
$\Delta R(j,\tau) \le 0.4$, and all jets with $\left| \eta^j \right| >
1.2$, are taken to have vanishing tagging probability. All other jets
with $E_{T}^j \geq 15$ GeV and $\left| \eta^j \right| \leq 1.2$ are assumed
to be (mis)tagged with probability $\eps_{u,d,s,g} = 0.01$.

Recall that we wish to identify events of the type $p \bar p
\rightarrow W^\pm h_2 \rightarrow W^\pm h_1 h_1 \rightarrow W^\pm b
\bar b b \bar b$. In order to avoid huge QCD backgrounds, and to
ensure that the event can be triggered on, we require the $W^\pm$ to
decay leptonically, $W^\pm \rightarrow \ell^\pm \nu_\ell$, with $\ell
= e$ or $\mu$. The neutrino will in general lead to sizable missing
transverse momentum, which helps to suppress backgrounds where a (real
or fake) lepton is produced from sources other than $W$ decay. We thus
apply the basic selection cuts:
\bea \label{tev-sel}
N_{\rm jet} &\ge& 4,\ {\rm with} \ E_{T}^{j=1-4} > 10 \ {\rm GeV \ and}, \
|\eta^{j=1-4}| < 3.0\,; \nonumber \\
N_{\rm lepton} &\ge& 1,\ {\rm with} \ E_{T}^{\ell} > 15 \ {\rm GeV \ and}, \
  |\eta^{\ell}| < 2.0\,; \nonumber \\
\met &>& 15 \ {\rm GeV}\,.
\eea

\begin{figure}[h!]
\begin{center}
\rotatebox{0}{\includegraphics[width=14cm]{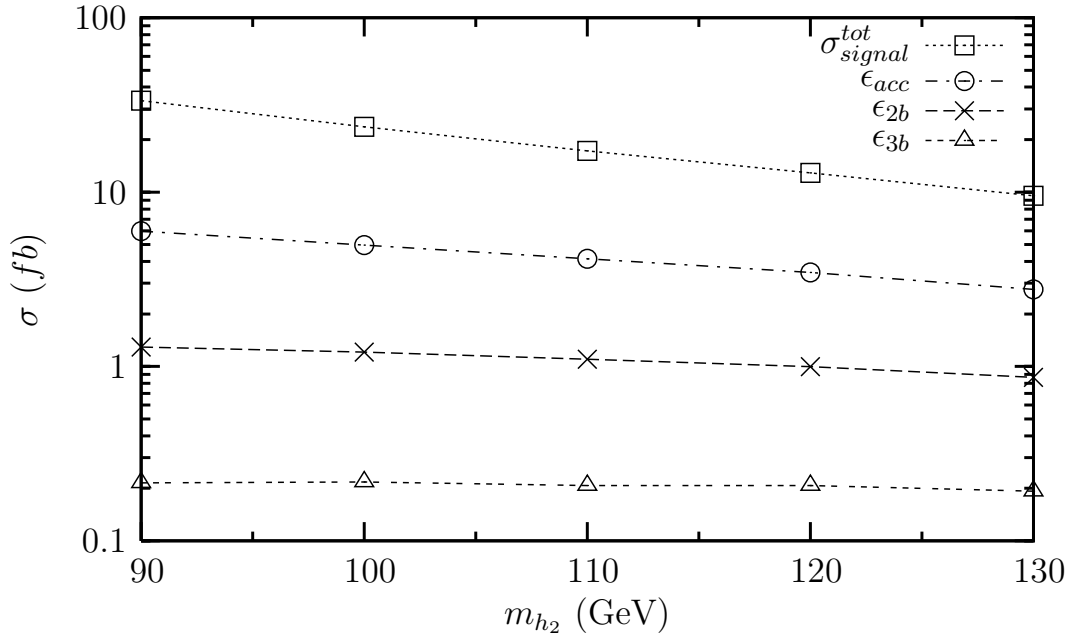}}
\caption{The $W\htwo$ signal cross section at the Tevatron with a 
leptonically decaying $W$. The four lines from the top to the bottom
correspond to, respectively, the raw cross section times branching
ratio, the same cross section after including the acceptance cuts
(\ref{tev-sel}), and the same cross section after double and triple
$b-$tagging.}
\label{crossTeV}
\end{center}
\end{figure}

Fig.~\ref{crossTeV} shows total signal cross sections at the Tevatron
as function of the mass of the heavier Higgs boson $h_2$. The
uppermost curve shows the total signal cross section times branching
ratio; it differs from the corresponding numbers in
Table~\ref{tab:bench} by a factor of 0.106. (Recall that we take
$C_{h_2WW} = 0.5$.) The dot--dashed curve shows the effect of imposing
the acceptance cuts (\ref{tev-sel}). We see that these simple cuts
reduce the signal cross section by a factor of about 6 (3.5) for
$m_{h_2} = 90 \ (130)$ GeV. The biggest reduction comes since we
require at least four jets in the final state. Especially for small
$m_{h_2}$ it is quite likely that some of the $b-$jets resulting from
$h_2$ decay will be too soft and/or too forward to be counted as
jets. In contrast, about 75\% of all signal events (with leptonically
decaying $W$ boson) contain one reconstructed lepton, and fully 92\%
of these events pass the mild $\met$ requirement in our selection
cuts.

Since we generated $W+4$ jets backgrounds with partonic $p_T$ down to
5 GeV, these background events are even less likely to have four
reconstructed jets. They also have slightly smaller efficiencies for
the leptonic and $\met$ requirements. However, in total the acceptance
cuts only increase the signal--to--background ratio by about a factor
of 2 for processes p2 through p8. Moreover, they actually favor
backgrounds p1, p9 and p10 containing a $t \bar t$ pair, which
contain much more energetic, central jets than the signal does.

\begin{table}[t!]
\begin{center}
\begin{tabular}{|c|r||c|c||c|c||c|c|}
\hline
& &\multicolumn{2}{|c||}{$N_{b} \ge 2 $} &
\multicolumn{2}{|c||}{ $N_{b} \ge 3 $ } & \multicolumn{2}{|c|}{$N_{b} \ge 4$} \\
Process & EvtSim & T & TM & T & TM &
T & TM\\
\hline 
%%%%%%%%%%%%%%%%%%%%%%%%%%%%%%%%%%%%%%%%%%%%%%%%%%%%%%%%%%%%%%%%%%%%%%%%%%%%%%%
S1 & 100000 & 0.2238 & 0.2250 & 0.0370 & 0.0374 & 0.00214 & 0.00222 \\
S2 & 100000 & 0.2051 & 0.2064 & 0.0309 & 0.0314 & 0.00160 & 0.00169 \\
S3 & 100000 & 0.1816 & 0.1827 & 0.0240 & 0.0244 & 0.00103 & 0.00108\\
S4 & 100000 & 0.1575 & 0.1587 & 0.0187 & 0.0191 & 0.00089 & 0.00099\\
S5 & 100000 & 0.1304 & 0.1315 & 0.0134 & 0.0137 & 0.00034 & 0.00037\\
CPX-1 & 100000 & 0.1566 & 0.1578 & 0.0183 & 0.0187 & 0.00076 & 0.00077\\
CPX-2 & 100000 & 0.1557 & 0.1568 & 0.0179 & 0.0183 & 0.00076 & 0.00080\\
\hline 
p1 & 2500000 & 0.1215 & 0.1402 & 0 & 0.0055 & 0 & 0.00007\\ 
p2 & 70000 & 0.0738 & 0.0742 & 0.0090 & 0.0090 & 0.00064 & 0.00066 \\ 
p3 & 125000 & 0.0393 & 0.0439 & 0.0022 & 0.0030 & 0 & 0.00004\\ 
p4 & 125000 & 0.0117 & 0.0195 & 0 & 0.0007 & 0 & 0.00002\\ 
p5 & 125000 & 0.0202 & 0.0289 & 0 & 0.0012 & 0 & 0.00003 \\ 
p6 & 150000 & 0.0198 & 0.0210 & 0 & 0.0002 & 0 & 0 \\ 
p7 & 250000 & 0 & 0.0047 & 0 & 0.00004 & 0 & 0 \\ 
p8 & 1540000 & 0 & 0.00011 & 0 & 0.0000011 & 0 & 0\\ 
p9 & 125000 & 0.2243 & 0.2402 & 0.0399 & 0.0485 & 0.00275 & 0.00426\\ 
p10 & 70000 & 0.1263 & 0.1677 & 0 & 0.0124 & 0 & 0.00053\\ 
%%%%%%%%%%%%%%%%%%%%%%%%%%%%%%%%%%%%%%%%%%%%%%%%%%%%%%%%%%%%%%%%%%%%%%%%%%%%%%%
\hline
\end{tabular}
\caption{The tagging efficiency without (T) and with the inclusion of
  mistagging (TM) efficiencies for signal and backgrounds at the
  Tevatron, for different minimal number of tagged $b-$jets per
  event. EvtSim stands for the number of events generated within the
  region of phase space defined by (\ref{presel}); no additional cuts
  have been applied.}
\label{tab:effTag_TeV}
\end{center}
\end{table}

We thus need some $b-$tagging to further suppress backgrounds.
Table~\ref{tab:procs} shows that we need to tag at least two $b-$jets
in order to reduce the a priori dominant background from process p8 to
a level comparable with or below the signal, given that we assume a
mistagging probability of 1\%. In Table~\ref{tab:effTag_TeV} we
therefore show the efficiency of requiring at least two, three or four
$b-$tags for our seven signal scenarios, as well as for the ten
classes of backgrounds. As described in Sec.~2 we have split process
p8 into numerous subclasses in order to improve the reliability of the
simulation; this explains the large number of events of this class we
generated. The tagging probabilities have been derived by counting
events which have the required minimum number of $b-$tags. Recall that
a jet has to be taggable, which in particular requires $|\eta| < 1.2$;
it is then tagged if a random number generated for this jet is less
than the (mis)tagging probability we assume for the relevant
flavor. Although this procedure is very similar to the treatment of
actual data, where an event would be either accepted or rejected, it
leads to relatively poor statistics if some or all of the $b-$tags are
actually mistags. In particular, for process p8 we can only state that
none of the generated events contained four tagged jets, leading to an
upper bound on the efficiency of order of $2 \cdot 10^{-6}$.

Nevertheless this Table clearly shows that increasing the number of
$b-$tags greatly increases the signal--to--background ratio. However, after
requiring at least two $b-$tags, processes p1, p6 and p8 still have
larger rates than the signal. $t \bar t$ events (p1) can be
efficiently reduced by imposing additional kinematical cuts, but the
jets in processes p2 through p8 have similar energies as those in the
signal. One then either has to look for invariant mass peaks to
isolate the signal on top of a sizable background, or further reduce
the background by increasing the number of $b-$tags.

Unfortunately Fig.~\ref{crossTeV} shows that the second of these
options would need several tens of fb$^{-1}$ of integrated luminosity
to generate a handful of signal events. This is due to the small
triple $b-$tagging probability, which lies between 1.3 and 3.6\%,
depending on $m_{h_2}$. Table~\ref{tab:eventTeV} shows that if this
luminosity was available, the signal would actually exceed the
background after imposing rather mild kinematical cuts. 

In addition to the selection cuts (\ref{tev-sel}) we demand that the
signal contains exactly (rather than at least) four jets. This reduces
combinatorial backgrounds for Higgs mass reconstructions. It also
reduces the signal as well as the $W+$ jets backgrounds (processes p2
through p7) by about 35\%, the inclusive $t \bar t$ background
(process p1) by about 60\%, and the $t \bar t Q \bar Q$ backgrounds
(processes p9 and p10) by about 85\%. Since the ``LEP hole'' requires
$m_{h_2} \lsim 130$ GeV, we next demand that the four--jet invariant
mass, which is the experimental estimator for $m_{h_2}$, is below 140
GeV. This reduces the signal only by 10 to 15\%, while reducing the
$W+$ jets backgrounds by about a factor of two. More importantly, it
reduces the inclusive $t \bar t$ background by about a factor of 150;
somewhat counter--intuitively, this cut reduces the $t \bar t Q \bar
Q$ backgrounds only by about a factor of 20, since the requirement of
having only four jets already required some of the partons to be
outside the acceptance region. It is important to notice that $t \bar
t$ production remains among the dominant backgrounds even after this
cut. In the absence of showering, the invariant mass of the four jets
in a $t \bar t$ event where one $t-$quark decays semi--leptonically
would have an invariant mass above $m_t$, well above the cut. However,
in some (small) fraction of events hard final state radiation carries
enough energy outside of the acceptance region to allow the event to
pass the cut; alternatively some jet(s) from top decay might be
outside of the acceptance region, with the missing jet(s) provided by
initial-- or final--state radiation. Since the cut value is well below
$m_t$, the jet energy resolution (\ref{smearing}) does not play a
major role here.

Finally, we pick the jet pairing $(ij) (kl)$ (with $i,j,k,l \in
\{1,2,3,4\}$) that minimizes the difference $|$ $m_{j_ij_j} - m_{j_kj_l}$ $|$
of di--jet invariant masses; in the absence of showering and for
perfect energy resolution, the signal would have $m_{j_ij_j} =
m_{j_kj_l} = m_{h_1}$. We then demand that both $m_{j_ij_j}$ and
$m_{j_kj_l}$ lie between 10 and 60 GeV, where the lower bound comes
from the requirement that $h_1 \rightarrow b \bar b$ decays should be
allowed, and the upper bound from the requirement that $h_2
\rightarrow h_1 h_1$ decays should be open. This very mild cut leads
to a modest further improvement of the signal to background ratio.

\begin{table}[t!] 
\begin{center}
\begin{tabular}{|c|c|c|c|c|c|c|}
\hline
Process& RawEvt & $N_{acc}$ & $N_{2b}$ & Eff2 (h2, +h1) & $N_{3b}$ &
Eff3 (h2, +h1) \\
\hline
%%%%%%%%%%%%%%%%%%%%%%%%%%%%%%%%%%%%%%%%%%%%%%%%%%%%%%%%%%%%%%%%%%%%%%%%%%%%%%%
S1 & 38.11 & 11.09 &  3.46 & 1.80(1.47,1.40) & 0.77 &  0.49 (0.42,0.40)\\
S2 & 51.59 & 13.85 &  3.99 & 2.07(1.76,1.69) & 0.83 &  0.51 (0.46,0.44)\\
S3 & 68.91 & 16.58 &  4.41 & 2.36(2.04,1.95) & 0.83 &  0.54 (0.48,0.47)\\
S4 & 94.76 & 19.88 &  4.85 & 2.55(2.21,2.11) & 0.87 &  0.54 (0.48,0.46)\\
S5 & 133.61   & 23.92 & 5.17 & 2.77(2.43,2.25) & 0.86 &0.52 (0.47,0.45)\\
CPX-1 & 89.89 & 20.27 & 4.72 & 2.48(2.18,2.08) & 0.82 &0.49 (0.44,0.43)\\
CPX-2 & 87.56 & 22.46 & 5.21 & 2.67(2.40,2.27) & 0.84 &0.53 (0.48,0.47)\\
\hline
p1 & 6760 & 3545 &599.5&191.7 ( 4.69, 3.86)  & 25.62 & 9.65 (0.06, 0.05)\\ 
p2 & 12.59 & 1.52 & 0.34 & 0.17 (0.09, 0.08) & 0.06 & 0.03 (0.01, 0.01)\\ 
p3 & 0.043 & 0.01 & 0 & 0 (0, 0) & 0 & 0 (0, 0) \\ 
p4 & 131.2 & 17.6 & 0.94 & 0.39 (0.22, 0.19) & 0.05 & 0.02 (0.01, 0.01) \\ 
p5 & 44.51 & 5.53 & 0.44 & 0.22 (0.12, 0.10) & 0.03 & 0.01 (0.01, 0.01)\\ 
p6 & 5181 & 610.2 & 34.71 & 16.19 (8.53, 7.31) & 0.28 & 0.15 (0.07, 0.06)\\ 
p7 & 14.31 & 2.01 & 0.02 & 0.01 (0, 0) & 0 & 0 (0, 0) \\ 
p8 & 384000 & 47340 & 14.65 & 4.44 (2.33, 2.03) & 0.24 & 0.02 (0 , 0)\\ 
p9 & 12.15 & 6.80 & 2.08 & 0.17 (0.01, 0.01) & 0.44 & 0.03(0, 0) \\ 
p10 &21.69 & 13.66 & 2.53 & 0.28 (0.01,0.01) & 0.19 & 0.02 (0, 0)\\ 
\hline
ToB & 396200 & 51550 & 655.2 & 213.6 (16.00, 13.59) & 26.91 & 9.93 (0.16,
0.14)
%%%%%%%%%%%%%%%%%%%%%%%%%%%%%%%%%%%%%%%%%%%%%%%%%%%%%%%%%%%%%%%%%%%%%%%%%%%%%%%
\\ \hline 
\end{tabular}
\caption{Expected number of events after different combinations of
  cuts for signal and backgrounds at the Tevatron with 4 fb$^{-1}$
  integrated luminosity. Expected event numbers below 0.005 have been
  given as 0. RawEvt stands for the number of events with only the
  generator--level cuts (\ref{presel}) imposed; for the signal as well
  as for background process p1, these are calculated from the total
  cross section times branching ratio. $N_{acc}$ is the number after
  the selection cuts (\ref{tev-sel}), whereas $N_{2b}$ and $N_{3b}$
  give the number of events with at least two or three jets tagged as
  $b-$jets, allowing for mistagging. The fifth column gives the
  number of events passing the selection cuts that contain exactly
  four jets, at least two of which are tagged as $b-$jets; the numbers
  in parentheses represent the number of events with the inclusion of
  dijet pair and four jet invariant mass cuts. The last column gives
  the number of events passing the selection cuts that contain exactly
  four jets, at least three of which are tagged as $b-$jets; the
  meaning of the numbers in parentheses is as in the fifth
  column. Finally, ToB is the total number of background events.}
\label{tab:eventTeV}
\end{center}
\end{table}

The results of this analysis are summarized in
Table~\ref{tab:eventTeV}. Recall from our discussion of
Table~\ref{tab:effTag_TeV} that many backgrounds have very small
triple $b-$tagging efficiencies, since some or all of these tags have
to be mistags. This leads to very poor statistics for these
backgrounds if events where not sufficiently many jets are tagged are
discarded. This method was used in deriving the numbers in the
$N_{2b}$ and $N_{3b}$ columns. The results of the column labeled Eff2
(Eff3) have instead been obtained by including all events that pass
the other cuts (not related to tagging) and have at least two (three)
{\em taggable} jets, assigning each event a weight given by its
(mis)tagging probability. This greatly increases the statistics. We
checked that this gives results that are consistent with the event
rejection technique whenever the latter has good statistics (see also
Table~\ref{tab:eventLHC} below); this is the case if at most one
$b-$tag results from mistagging.

\begin{figure}[h!]
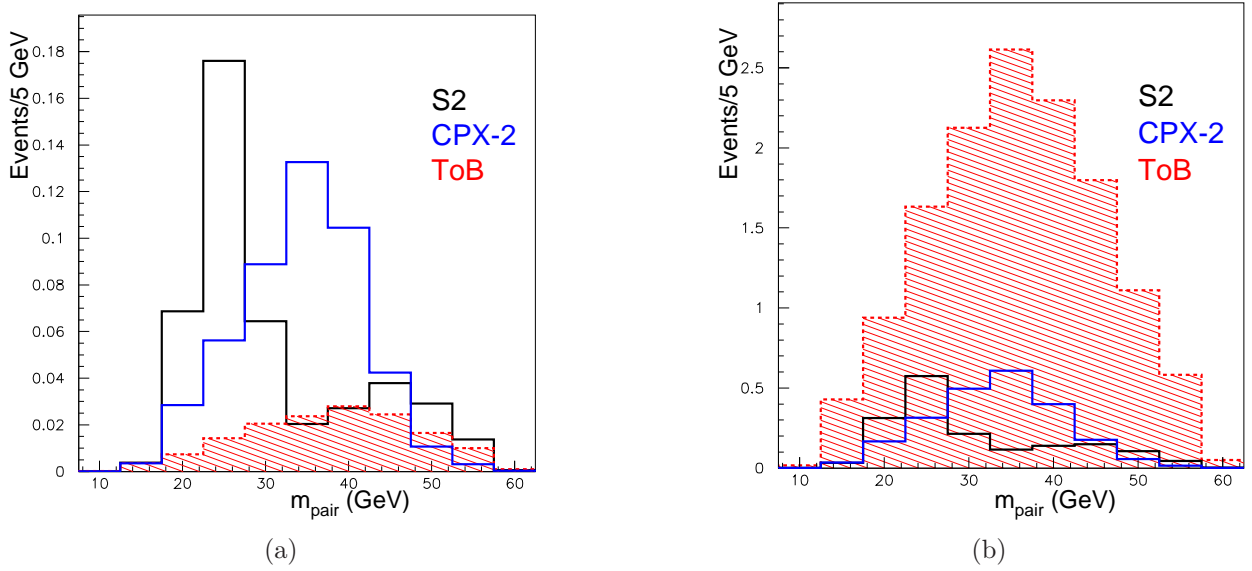

\setlength{\unitlength}{1.2in}
\subfigure[\label{tev_mh1_3bl}]{
    \begin{picture}(3,2.3)
      \put(0.3,0){\epsfig{file=tev_h1_3b.epsi, width=2.75in}}
    \end{picture}
  }%\hfill  commenting for less space between two subfigs.
  \subfigure[
	\label{tev_mh1_2bl}]{
    \begin{picture}(3,2.3)
      \put(0.3,0){\epsfig{file=tev_h1_2b.epsi, width=2.75in}}
    \end{picture}
  }
  \caption{The average di--jet invariant mass $m_{\rm pair}$ defined
    in Eq.(\ref{h1}) for the S2 and CPX-2 signal benchmark points
    after all cuts compared with total background (ToB), for either
    triple (left) or double (right) $b-$tag. In the absence of
    showering and energy smearing, $m_{\rm pair} = m_{h_1}$ for the
    signal. The distributions for the other signal points S$i$ are
    very similar to S2, since they all have $\mhone=30$ GeV; the
    distribution for CPX-1 peaks near 30 GeV. The left (right)
    distributions have been obtained using the event weighting
    technique, as in the 7th (5th) column of
    Table.~\ref{tab:eventTeV}.}

\label{tev_mh1}
\end{figure}

As mentioned above, in the absence of showering and energy smearing
and for stable $b-$quarks, in signal events it should be possible to
form two jet pairs out of the four jets such that $m_{j_ij_j} =
m_{j_kj_l} = m_{h_1}$. We attempt to reconstruct the $h_1$ mass using
the optimal jet pairing found above, and defining
\beq \label{h1}
m_{\rm pair} = \frac {1}{2} \left( m_{j_ij_j} + m_{j_kj_l} \right)\,.
\eeq
By averaging the two jet--pair invariant masses in a given event, we
reduce fluctuations. The left frame in Fig.~\ref{tev_mh1} shows the
distribution of this variable for signal scenarios S2 and CPX-2 as
well as for the total background. We see clear peaks in the signal on
top of a background that is shaped by the requirement that {\em both}
di--jet invariant masses should lie between 10 and 60 GeV. The peaks
are somewhat below $m_{h_1}$, partly due to the relatively small jet
cone size $R(j) = 0.4$ we use, and partly because most signal events
contain neutrinos from semi--leptonic $b$ or $c$ decays. (The charm
quarks themselves are produced in $b$ decays.)

\begin{figure*}[h!]
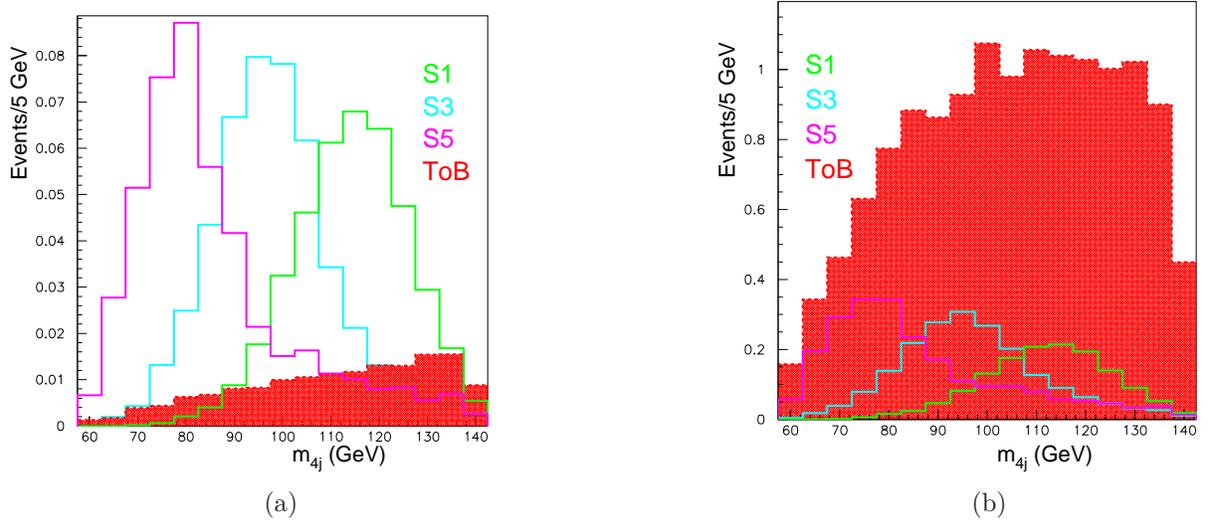

\setlength{\unitlength}{1.2in}
\subfigure[]{
    \begin{picture}(3,2.3)
      \put(0.3,0){\epsfig{file=tev_h2_3b.epsi, width=2.5in}}
    \end{picture}
  }%\hfill  commenting for less space between two subfigs.
  \subfigure[]{
    \begin{picture}(3,2.3)
      \put(0.3,0){\epsfig{file=tev_h2_2b.epsi, width=2.5in}}
    \end{picture}
  }
\caption{The four--jet invariant mass $m_{4j}$ distribution after all
  cuts for signal scenarios S1, S3 and S5 and for the total background
  (ToB), requiring triple (left) our double (right) $b-$tag. 
   The left (right) distributions have been obtained using the event weighting
   technique, as in the 7th (5th) column of Table.~\ref{tab:eventTeV}.}
\label{tev_mh2} 
\end{figure*}

In the absence of showering etc. the signal should have fixed
four--jet invariant mass $m_{4j}$ equal to $m_{h_2}$. The distribution
of this variable is shown in the left frame of Fig.~\ref{tev_mh2} for
signal scenarios S1, S3 and S5 as well as for the total background. We
again observe clear peaks for the signal, again shifted downwards (by
10 to 15 GeV) from the naive expectation $m_{4j} = m_{h_2}$.

The peaks in the $m_{\rm pair}$ and $m_{4j}$ distributions allow to
define the final significance of the signal by counting events that
satisfy 
\bea \label{doublepeak}
0.6 m_{h_1} &\leq m_{\rm pair} &\leq m_{h_1} + 5 \ {\rm GeV}\,;
\nonumber \\
0.7 m_{h_2} &\leq m_{4j} &\leq m_{h_2} + 10 \ {\rm GeV}\,.
\eea
The resulting significances, calculated as $S/\sqrt{B}$ for a total
integrated luminosity of 20 fb$^{-1}$, are tabulated in
Table~\ref{tab:TeV_signi}. This integrated luminosity now seems within
reach after including results from both experiments. The significance
defined in this way overestimates the true statistical significance of
a double peak in the $m_{\rm pair}$ and $m_{4j}$ distributions
somewhat, due to the ``look elsewhere'' effect: since $m_{h_1}$ and
$m_{h_2}$ are not known a priori, one would need to try different
combinations when looking for peaks. However, given that we use rather
broad search windows, there are probably only ${\cal O}(10)$
statistically independent combinations within the limits of the LEP
hole.

\begin{table}[t!] 
\begin{center}
\begin{tabular}{|c|c|c|c|c|c|c|}
\hline
 & \multicolumn{3}{c|}{$N_b \geq 3$} & \multicolumn{3}{c|}{$N_b \geq 2$} \\
Scenario & $S$ & $B$ & ${\cal S}$ & $S$ & $B$ & ${\cal S}$ \\
\hline
%%%%%%%%%%%%%%%%%%%%%%%%%%%%%%%%%%%%%%%%%%%%%%%%%%%
S1 & 1.49 & 0.14 & 3.98 & 4.78 & 15.48 & 1.21 \\
S2 & 1.51 & 0.15 & 3.89 & 5.25 & 16.61 & 1.28 \\
S3 & 1.47 & 0.15 & 3.79 & 5.71 & 18.13 & 1.34 \\
S4 & 1.54 & 0.14 & 4.12 & 6.45 & 18.61 & 1.49 \\
S5 & 1.56 & 0.13 & 4.33 & 7.25 & 17.15 & 1.75 \\
CPX-1 & 1.62 & 0.22 & 3.45 & 7.29 & 26.22 & 1.42 \\
CPX-2 & 1.69 & 0.25 & 3.38 &  7.24 & 27.75 & 1.37\\
%%%%%%%%%%%%%%%%%%%%%%%%%%%%%%%%%%%%%%%%%%%%%%%%%%%
\hline
\end{tabular}
\caption{Final number of signal $S$ and background $B$ events and the
  corresponding significance ${\cal S}$ at the Tevatron, defined as
  ${\cal S} = S / \sqrt{B}$. We have assumed an integrated luminosity of
  20 fb$^{-1}$, and applied all cuts, including the double peak
  requirement (\ref{doublepeak}). We show results separately requiring
  at least two or at least three tagged $b-$jets.}
\label{tab:TeV_signi}
\end{center}
\end{table}

We see that requiring triple $b-$tags leads to very good
signal to background ratio, of around 10 for $m_{h_1} = 30$ GeV and
slightly less for heavier $h_1$. However, we expect less than 2 signal
events after all cuts even in the assumed large data sample. The
nominal significance exceeds three, but of course Gaussian statistics
is not appropriate for these small event numbers. For example, for
scenario S2 in the absence of a signal the probability to see no event
after cuts is about 86\%, but the probability for finding one event is
13\%, and that for finding two events is about 1\%. After adding the
signal, the probability for observing zero or one event is about 53\%,
while the probability of finding three or more events is only 21\%. We
conclude that an analysis requiring triple $b-$tag will probably not
lead to a significant signal.

We saw that the problem is the low number of events left after all
cuts, which is partly due to the poor efficiency of the
signal. Clearly we need at least four jets in order to be able to
reconstruct $m_{h_2}$, which in turn is crucial for the final double
peak analysis. The cuts on the missing $E_T$ and the leptonic $p_T$
are already quite mild. The only cut one may relax is thus the
requirement of triple $b-$tag. We see in Table~\ref{tab:eventTeV} that
reducing the number of $b-$tags to two increases the signal rate by a
factor between 3.7 and 5. Unfortunately it also increases the total
background by two orders of magnitude, the main sources being events
with two real $b-$quarks in the final state (processes p1 and p6), but
background class p8, without real $b$ in the final state, now also
contributes significantly. The right frames in Figs.~\ref{tev_mh1} and
\ref{tev_mh2} show that the peaks in the di--jet and four--jet
invariant masses are now buried in the background. Not surprisingly,
Table~\ref{tab:TeV_signi} finds a statistical significance of well
below two if only two $b$ tags are required. 

Note also that the signal rate is still quite small. Further
kinematical cuts, which might slightly increase the signal to
background ratio, are therefore not likely to increase the statistical
significance of the signal. We are therefore forced to conclude that
the search for $W h_2 \rightarrow W h_1 h_1 \rightarrow \ell \nu b
\bar b b \bar b$ events at the Tevatron does not seem promising, and
turn instead to the LHC.

\section{LHC}
\label{lhc_analysis}

Our analysis for the LHC follows broadly similar lines as that for the
Tevatron. However, there are significant quantitative differences. On
the one hand, we expect improved detector performance and a higher
integrated luminosity at the LHC. On the other hand, we saw in Sec.~2
that increasing the beam energy and going from $pp$ to $p \bar p$
collisions reduces the signal to background ratio before cuts by about
one order of magnitude.

We simulate our signal and backgrounds at the LHC with $\sqrt s = 14$
TeV. The {\tt PYCELL} model is based on the ATLAS detector \cite{atlas}.
Specifically, we assume calorimeter coverage $\rm |\eta| < 5.0$, with
segmentation $\Delta \eta \times \Delta \phi = 0.087 \times 0.10$. We
again use the same Gaussian energy resolution for leptons and jets,
with
\beq \label{lhc_res}
{\Delta E^{j,\ell} \over E^{j,\ell}} = {50\% \over \sqrt{E^{j,\ell}}}
\oplus 3\% \quad . 
\eeq

As before, we use a cone algorithm for jet finding, with jet radius
$\rm\Delta R(j) = \sqrt{\Delta\eta^{2}+\Delta\phi^{2}} = 0.4$.
Calorimeter cells with $\rm E_{T,min}^{cell} \ge 1.0$ GeV are considered
to be potential candidates for jet initiator. All cells with $\rm
E_{T,min}^{cell} \ge 0.1$ GeV are treated as part of the would--be jet. A
jet is required to have minimum summed $\rm E_{T,min}^{jet} \ge 15$
GeV.

Leptons ($\rm \ell = e, ~\mu$) are selected if they satisfy $\rm E_T^{\ell} 
\ge 20$ GeV and $\rm |\eta^{\ell}| \le 2.5$. The jet--lepton isolation
criterion is as in the Tevatron analysis. The missing transverse
energy $\met$ is also determined in the same way as at the Tevatron
(however with better angular coverage of the calorimeter, as described
above).

Only jets with $|\eta^j| < 2.5$ are considered to be taggable as $b-$jets. If
the jet is ``matched'' to a $b-$flavored hadron, with $\Delta R(j,{\rm
  hadron}) \leq 0.2$, the tagging efficiency is taken to be 50\%. If instead
the jet is matched to a $c-$hadron, the (mis)tagging efficiency is taken to be
10\%, whereas jets matched to a $\tau-$lepton have zero tagging
probability. All other taggable jets have (mis)tagging probability of
0.25\%. These efficiencies follow recent ATLAS and CMS
analyses~\cite{Aad:2009wy,Lehmacher:2008hs,mistaggcharm}.

We then apply the following basic selection cuts:
\bea \label{lhc-sel}
N_{\rm jet} &\ge& 4,\ {\rm with} \ E_{T}^{j=1-4} > 15 \ {\rm GeV \ and}, \
|\eta^{j=1-4}| < 5.0\,; \nonumber \\
N_{\rm lepton} &\ge& 1,\ {\rm with} \ E_{T}^{\ell} > 20 \ {\rm GeV \ and}, \
  |\eta^{\ell}| < 2.5\,; \nonumber \\
\met &>& 20 \ {\rm GeV}\,.
\eea

\begin{figure}[h!]
\begin{center}
\rotatebox{0}{\includegraphics[width=14cm]{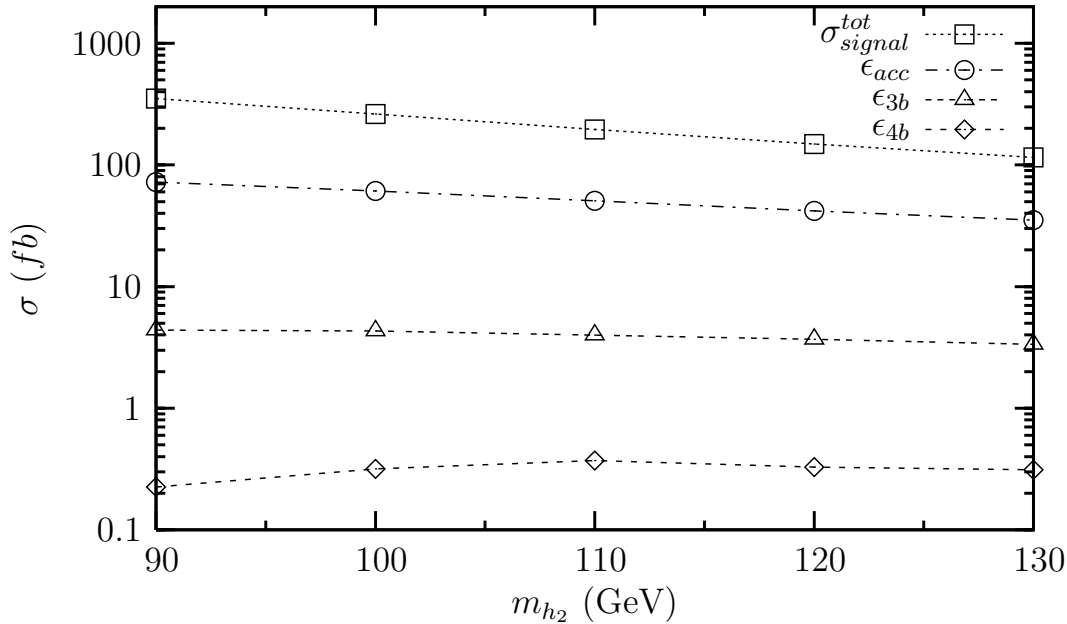}}
\caption{The signal cross section as function of $m_{h_2}$ at the LHC.
  The four lines from the top to the bottom correspond to the total
  cross section times branching ratio, the cross section passing the
  selection cuts (\ref{lhc-sel}), and the cross sections requiring at
  least three and four tagged $b-$jets, respectively.}
\label{crossLHC}
\end{center}
\end{figure}

Fig.~\ref{crossLHC} shows that these cuts reduce the cross section by
about a factor of 5 (3) for $m_{h_2} = 90 \ (130)$ GeV. Comparison
with Fig.~\ref{crossTeV} shows that the signal efficiency is slightly
higher at the LHC. This is due to the higher probability to find four
jets in the event, partly due to the better calorimeter coverage, and
partly because of increased showering at the higher LHC energy. In
contrast, the increased thresholds for $E_T^\ell$ and $\met$ slightly
reduce the efficiencies of these cuts compared to the Tevatron analysis.

\begin{table}[t!]
\begin{center}  
\begin{tabular}{|r|r|c|c|c|c|c|}
\hline
Process & RawEvt & $N_{acc}$ & $N_{3b}$ & $N_{4b}$ & Eff3C (h2, +h1) &
Eff3T (h2, +h1)\\
\hline
%%%%%%%%%%%%%%%%%%%%%%%%%%%%%%%%%%%%%%%%%%%%%%%%%%%%%%%%%%%%%%%%%%%%%%%%%%%%%%%
S1 &1156.52&352.50&33.48&3.12&13.72(6.82,6.46)&13.96(7.30,6.85)\\
S2 &1485.59&418.27&36.84&3.28&15.05(7.37,6.89)&15.36(8.30,7.76)\\
S3 &1961.82&506.54&39.81&3.71&17.05(9.06,8.61)&17.03(9.45,8.91)\\
S4 &2620.32&610.64&43.18&3.17&18.94(10.61,10.06)&17.81(10.16,9.61)\\
S5 &3516.41&724.70&43.92&2.25&19.09(9.28,8.86)&18.96(10.15,9.63)\\
CPX-1&2509.17&600.19&40.07&2.71&16.54(8.93,8.28)&17.25(9.57,9.07)\\
CPX-2&2420.86&597.18&40.28&2.78&17.16(9.78,9.25)&16.88(10.11,9.64)\\
\hline
p1 & 1,690,000 & 818,800 & 7795 & 111.0 & 1558 (7.94, 6.08) & 1469(7.57, 5.52)\\
p2 & 337.6 & 31.8 & 4.10 & 0.46 & 3.07 (0.63, 0.54) & 2.95 (0.63, 0.53)\\ 
p3 & 23.3 & 2.3 & 0.13 & 0.01 & 0.10 (0.01, 0.01) & 0.11 (0.02, 0.01)\\ 
p4 & 73,170 & 7359 & 77.56 & 0.59 & 55.32 (8.20, 7.32) & 56.50 (7.90, 6.79)\\ 
p5 & 1126 & 89.9 & 1.68 & 0.05 & 1.22 (0.32, 0.27) & 1.17 (0.28, 0.25)\\ 
p6 & 535,700 & 45,830 & 17.14 & 0 & 8.57 (0, 0) & 17.89 (2.25, 1.93)\\ 
p7 & 7194 & 586.3 & 0.23 & 0 & 0.17 (0.06, 0.06) & 0.05 (0.01, 0.01)\\ 
p8 & 59,700,000 & 4,332,000 & 2.18 & 0 & 1.35 (0.01, 0.01) & 4.59
(0.75, 0.68) \\ 
p9 & 10,100 & 5700 & 751.5 & 96.26 & 78.56 (1.45, 1.21) & 72.82 (1.49,
1.28)\\ 
p10 & 16,440 & 9245 & 259.8 & 11.18 & 35.76 (0, 0) & 31.54 (0.53, 0.45)\\ 
\hline
ToB & 62,030,000 & 5,220,000 & 8910 & 219.6 & 1742 (18.62, 15.50) &
1657 (21.43, 17.45)\\ 
%%%%%%%%%%%%%%%%%%%%%%%%%%%%%%%%%%%%%%%%%%%%%%%%%%%%%%%%%%%%%%%%%%%%%%%%%%%%%%%
\hline 
\end{tabular}
\caption{Expected number of events after different combinations of
  cuts for signal and backgrounds at the LHC with 10 fb$^{-1}$
  integrated luminosity. The notation is similar to that of
  Table~\ref{tab:eventTeV}, but we do not show results for only
  double $b-$tag, and instead show the expected number of events with
  four (or more) $b-$tags. However, the last two columns refer to
  final efficiencies requiring at least three $b-$tags, as explained
  in the text.}
\label{tab:eventLHC}
\end{center}
\end{table}  

Due to the reduced raw signal to background ratio, at the LHC one will
definitely have to require at least three $b-$tags in each
event. Fig.~\ref{crossLHC} and Table~\ref{tab:eventLHC} show
that requiring a fourth $b-$tag reduces the signal cross section
by another order of magnitude or more. The signal rate then becomes
so low that one would have to wait for the high--luminosity phase of
the LHC to accumulate enough events to reconstruct invariant mass
peaks. However, in that phase the $b-$tagging performance might be
degraded, since then ${\cal O}(20)$ $pp$ collisions will occur in a
single bunch crossing. We therefore stick to triple $b-$tag in our LHC
analysis. 

Note that the total $b-$tagging efficiency of signal events is
somewhat higher at the LHC than at the Tevatron. For example, for
scenario S2 we find that 8.6\% of all signal events that pass the
basic acceptance cuts (\ref{lhc-sel}) contain at least three $b-$tags,
compared to 6.0\% at the Tevatron (see Table~\ref{tab:eventTeV}). This
is mostly due to the larger rapidity coverage of the ATLAS vertex
detector. As at the Tevatron, the $b-$tagging efficiency of signal
events increases with $m_{h_2}$, but is largely independent of
$m_{h_1}$. As a result, the number of signal events containing three
or more $b-$tags is quite similar for all scenarios we consider.

Table~\ref{tab:eventLHC} also shows the impact of requiring at least
three or four $b-$tags on the background processes we consider. We see
that background processes containing less than two $b-$quarks are
suppressed to a level well below the signal by the triple $b-$tag
requirement. This is true in particular for background p8, which had
the largest cross section prior to $b-$tagging; after the triple
$b-$tag, this class of backgrounds is dominated by subclass p8.7,
which has three charm quarks in the final state.\footnote{Subclass
  p8.8, with four charm quarks in the final state, has higher tagging
  efficiency but much smaller total cross section. Conversely,
  subclass p8.6 with two charm quarks in the final state has two times
  larger total cross section but greatly reduced tagging
  probability. Note that we generated comparable numbers of events for
  all subclasses of p8, even though they have very different total
  cross sections.} Backgrounds p4 and p6, which contain exactly two
$b-$quarks in the final state, become about two times larger than and
comparable to the signal, respectively, after triple
$b-$tagging. Inclusive $t \bar t$ production still exceeds the signal
by more than two orders of magnitude, with about 10\% of this
background coming from classes p9 and p10 which have an additional
heavy quark pair in the final state. The total background still
exceeds the signal by a factor of 200 even after requiring three
$b-$tags.

We thus need to apply further kinematical cuts. To this end, and also
to show the basic event characteristics, we show some normalized
kinematical distributions of signal and backgrounds at the LHC. The
shapes of these distributions is actually rather similar at the
Tevatron; however, we saw that the number of events with three or more
tagged $b-$jets is too small to allow a meaningful measurement of such
distributions.

\begin{figure*}[h!]
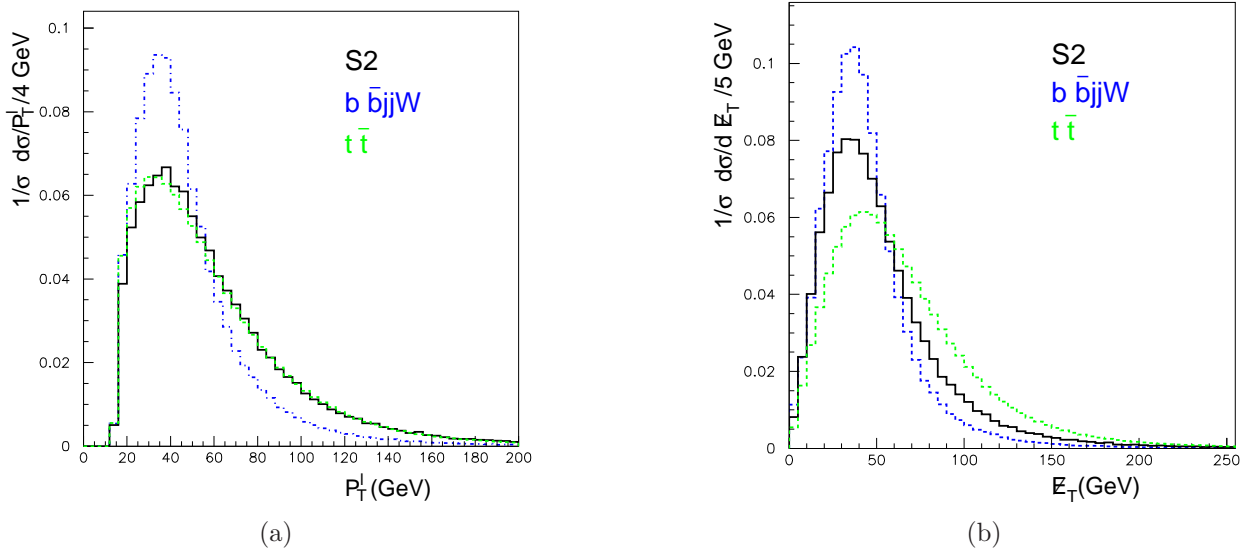

  \setlength{\unitlength}{1.2in}
  \subfigure[
	\label{lhc_lep_pt}]{
    \begin{picture}(3,2.3)
      \put(0.3,0){\epsfig{file=lhc_ptlep.epsi, width=2.75in}}
    \end{picture}
  }%\hfill  commenting for less space between two subfigs.
  \subfigure[
	\label{lhc_met}]{
    \begin{picture}(3,2.3)
      \put(0.3,0){\epsfig{file=lhc_met.epsi, width=2.75in}}
    \end{picture}
  }
  \caption{Normalized distributions of the lepton's transverse
    momentum (left) and total missing transverse energy ($\met$) (right)
    at the LHC. The solid black histograms are for the signal in scenario
    S2, while the dashed blue (dark grey) and green (light grey)
    histograms are for background processes p6 and p1,
    respectively. The results for processes p9 and p10 are similar to
    that for $t \bar t$ (p1); all the other backgrounds are similar to
    $b \bar b jjW$ (p6) as in all these cases the lepton and hard
    neutrino (giving $\met$) is from only one $W$ decay. Note that
    only the generator--level cuts (\ref{presel}) have been imposed.}
\label{lhc_lep_met}
\end{figure*} 

The left frame in Fig.~\ref{lhc_lep_met} shows the normalized $p_T$
distribution of the charged lepton in the event. The signal (black)
features a spectrum that is harder than that of the $W+4j$
backgrounds, represented by process p6 (blue or dark grey), and
similar to that of the $t \bar t$ backgrounds, represented by process
p1 (green or light grey). In the signal the $W$ recoils against a
single massive particle $h_2$, giving it a rather large transverse
momentum on average. In contrast, the transverse momenta of the four
jets in the $W+4j$ backgrounds will on average only add quadratically,
explaining the softer spectrum. On the other hand, in the $t \bar t$
backgrounds the leptonically decaying $W$ boson itself results from
the decay of one of the massive $t$ quarks, also giving it a typically
quite large transverse momentum.

Similar remarks apply to the missing $E_T$ distributions shown in the
right frame of Fig.~\ref{lhc_lep_met}. However, these distributions
peak at somewhat larger values, and have longer tails, than the
leptonic $p_T$ distributions. In case of the signal and the $W+4j$
backgrounds this is partly due measurement errors on the jets
contributing to the measured $\met$, and partly due to additional
softer neutrinos from semi--leptonic $b$ and $c$ decays. The $t \bar
t$ backgrounds can in addition have a hard neutrino coming from the
semi--leptonic decay of the second top quark. As a result, $t \bar t$
production features the hardest $\met$ spectrum of all the processes
we consider.

Figs.~\ref{lhc_lep_met} indicate that we could slightly increase the
ratio of signal to $W+4j$ backgrounds by increasing the cut values on
$p_T^\ell$ and/or $\met$. However, a harder $\met$ cut would reduce
the ratio of signal to $t \bar t$ backgrounds. Worse, either cut would
significantly reduce the signal rate, which in any case is not very
large. We conclude that changes of the $p_T^\ell$ or $\met$ cuts are
not likely to significantly improve the observability of our signal.

\begin{figure*}[h!]
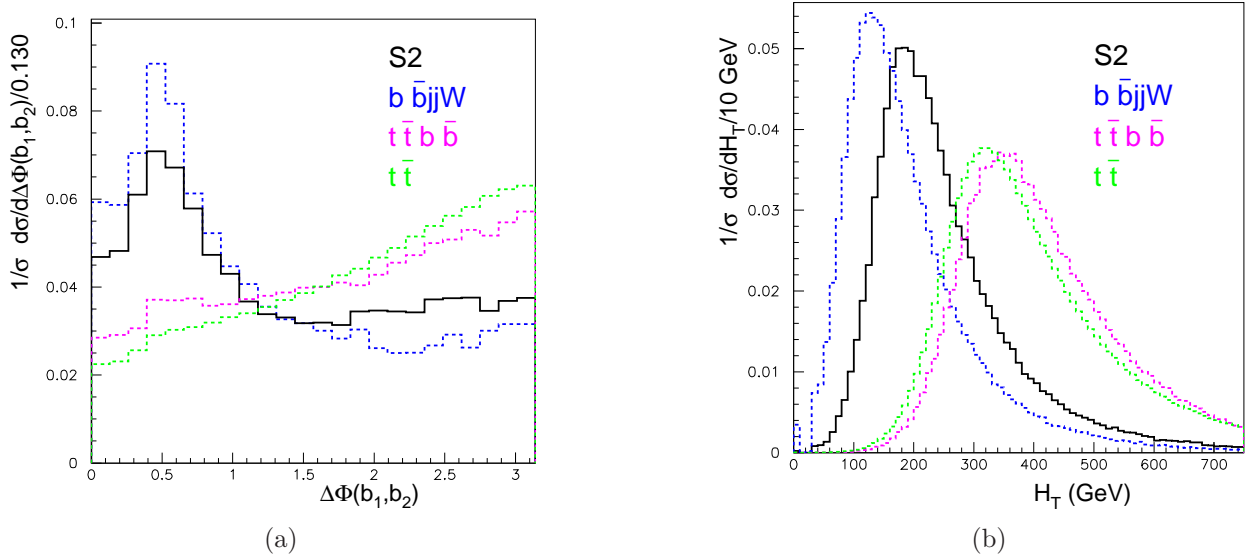

  \setlength{\unitlength}{1.2in}
  \subfigure[
	\label{lhc_delta_phi_tag_mistag}]{
    \begin{picture}(3,2.3)
      \put(0.3,0){\epsfig{file=lhc_delphi.epsi, width=2.75in}}
    \end{picture}
  }%\hfill  commenting for less space between two subfigs.
  \subfigure[
	\label{lhc_ht}]{
    \begin{picture}(3,2.3)
      \put(0.3,0){\epsfig{file=lhc_ht.epsi, width=2.75in}}
    \end{picture}
  }
\caption{Normalized distribution of (left) the opening angle $\Delta\Phi
  (b_1,b_2)$ between the two hardest tagged jets, allowing for mistagging,
  (right) the total hard scalar transverse energy $H_T$ defined in
  eq.(\ref{ht}). Black solid histograms are for the signal, while the blue
  (dark grey), magenta (grey) and green (light grey) dashed histograms are for
  background processes p6, p9 and p1, respectively. In both cases the
  distributions for $t \bar t c \bar c$ (p10) are similar to $t \bar t b \bar
  b$ (p9), while the other $W+4j$ backgrounds look similar to p6, although the
  height of the peak of the $\Delta \Phi(b_1,b_2)$ distribution differs
  somewhat for the different backgrounds. In the left panel, only events
  containing at least two tagged jets contribute.}
  \label{lhc_delta_phi_ht}
\end{figure*}

The left frame of Fig.~\ref{lhc_delta_phi_ht} shows the normalized
distribution in the opening angle between the two hardest jets that
have been tagged as $b-$jets, allowing for mistagging; evidently only
events containing at least two $b-$tags contribute. The notation is as
in Fig.~\ref{lhc_lep_met}, except that we in addition show results for
the $t \bar t b \bar b$ background (process p9). We see that the
signal has a clear peak at $\Delta \Phi(b_1,b_2) \simeq 0.5$,
corresponding to an opening angle of about $30^\circ$. This is because
the two most energetic $b-$jets tend to come from the decay of the
same $h_1$ boson, which is quite energetic, giving a sizable boost to
the $b-$quarks when going from the $h_1$ rest frame to the lab
frame. In contrast, $t \bar t$ backgrounds tend to have the two
leading $b-$jets in opposite hemispheres, since they come from the $t$
and $\bar t$ quark which recoil against each other. On the other hand,
in the $W+4j$ backgrounds the two leading $b-$jets tend to be even
closer together than in the signal, since this reduces the invariant
mass of the corresponding $b \bar b$ pair, and hence the virtuality of
the gluon from which it originated.

The right frame of Fig.~\ref{lhc_delta_phi_ht} depicts the
distribution of the total hard transverse energy $H_T$, defined as
\beq \label{ht}
H_T = \met + \sum_{j,\ell} E_T\,.
\eeq
We see that, as in the right frame of Fig.~\ref{lhc_lep_met}, the
distribution of the signal is harder than that of the $W+4j$
backgrounds, but softer than that of $t \bar t$ events. Hence a cut on
either $\Delta \Phi (b_1, b_2)$ or $H_T$ could enhance the signal
relative to one class of backgrounds, but would favor the other class
of backgrounds even more. Moreover, a significant increase of the
ratio of the signal to one class of backgrounds could only be achieved
at the cost of a sizable reduction of the signal. Once again, cutting
on these variables is not likely to yield a sizable increase of the
significance of the signal.

This leaves us with cuts on invariant masses, as we already employed
at the Tevatron. To that end, we again require the signal to have
exactly four reconstructed jets. This reduces the signal by slightly
more than a factor of two. This cut is more severe at the LHC due to
the much larger available phase space, and also due to the better
coverage of the calorimeter which is able to detect jets at quite
small angles. This cut reduces $W+4j$ backgrounds by slightly less
than a factor of two, since we used a smaller shower scale to account
for the fact that some of the ``hard'' jets are typically already
quite soft in these backgrounds. On the other hand, about 75\% of all
$t \bar t$ events passing the acceptance cuts (\ref{lhc-sel}) have at
least a fifth jet. 

The number of events (in 10 fb$^{-1}$ of integrated luminosity)
passing the acceptance cuts and containing exactly four jets, at least
three of which are tagged, is given by Eff3C and Eff3T in the last two
columns of Table~\ref{tab:eventLHC}. These columns differ in the way
the cross sections have been estimated. In the Eff3C column we have
discarded all events not containing at least three tagged jets, where
each taggable jet is tagged with the appropriate probability; this
closely mimics how a measurement of this cross section would be
performed. In contrast, in the Eff3T column we have counted all events
containing at least three taggable jets, but weighted them with the
appropriate tagging probability. As explained in the Tevatron Section,
this increases the statistics, and thus reduces the statistical
uncertainty; this is true in particular if one or more tags have to be
mistags. We therefore consider the estimate Eff3T to be more
reliable. It is reassuring to see that the two estimates agree quite
well not only for the signal, but also for those background that
(often) contain at least one $c$ quark in addition to a $b \bar b$
pair, which is true for both p1 and p4. The difference between the two
estimates becomes large only if the overall tagging efficiency is very
poor, as in background p6 (where at least one light flavor or gluon
jet has to be mistagged) and p8 (where all three tags are mistags,
although typically of $c$ jets, as noted above). In our Tevatron
analysis we had therefore only shown results using this latter
estimate.

We next require the four--jet invariant mass to lie between 60 and 140
GeV. Recall that at the parton level this invariant mass should be
equal to $m_{h_2}$ for the signal, so that this requirement covers the
entire ``LEP hole'' in the MSSM Higgs parameter space. The effect of
this cut is given by the first number in parentheses in the last two
columns of Table~\ref{tab:eventLHC}. The requirement $m_{4j} > 60$ GeV
reduces some of the $W+4j$ backgrounds significantly. More
importantly, the requirement $m_{4j} < 140$ GeV reduces the inclusive
$t \bar t$ background by about a factor of 200, and the $t \bar t Q
\bar Q$ backgrounds ($Q = b$ or $c$) by a factor of 50; it also
further reduces the $W+4j$ backgrounds. This is quite similar to the
situation at the Tevatron, see Table~\ref{tab:eventTeV}. Unfortunately
the four--jet invariant mass cut also reduces the signal by nearly a
factor of 2. The reason is that frequently one of the four $b-$quarks
is too soft to be counted as a jet. The fourth jet is instead provided
by initial state radiation. This allows four--jet invariant masses
well above $m_{h_2}$. The loss of signal is larger than at the
Tevatron, where the $E_T$ threshold for jets was taken to be 10 GeV,
rather than 15 GeV at the LHC; also, there is significantly more
radiation at the LHC.

Finally, we determine the optimal jet pairing by minimizing the
difference between the di--jet invariant masses, and require both of
these jet pair invariant masses to lie between 10 and 60 GeV. This
cut results in the last number in parentheses in the last two columns
of Table~\ref{tab:eventLHC}. As at the Tevatron, the impact of this
cut is rather mild for the signal and somewhat more pronounced for the
background, in particular that involving $t \bar t$ production.

After these cuts we are left with slightly less than one signal event
and slightly less than two background events per fb$^{-1}$ of data. A
$5 \sigma$ signal would then require almost 100 fb$^{-1}$ of data,
more than the LHC is likely to collect during ``low'' luminosity
running. Besides, the background prediction also has considerable
systematic uncertainties. The biggest background after all cuts comes
from $t \bar t$ production (p1$+$p9$+$p10), and depends sensitively on
the modeling of the four--jet invariant mass distribution. This in
turn depends not only on a correct treatment of radiative processes,
without which this background could not contribute at all; it would
also be affected significantly by a permille--level jet reconstruction
inefficiency within the nominal acceptance region of the calorimeter,
which could increase the probability that one of the partons from top
decay escapes detection. Moreover, the $W+4j$ cross sections have been
calculated in leading order QCD, and thus suffer from large scale
uncertainties.

\begin{figure}[h!]
\begin{center}
\rotatebox{0}{\includegraphics[width=14cm]{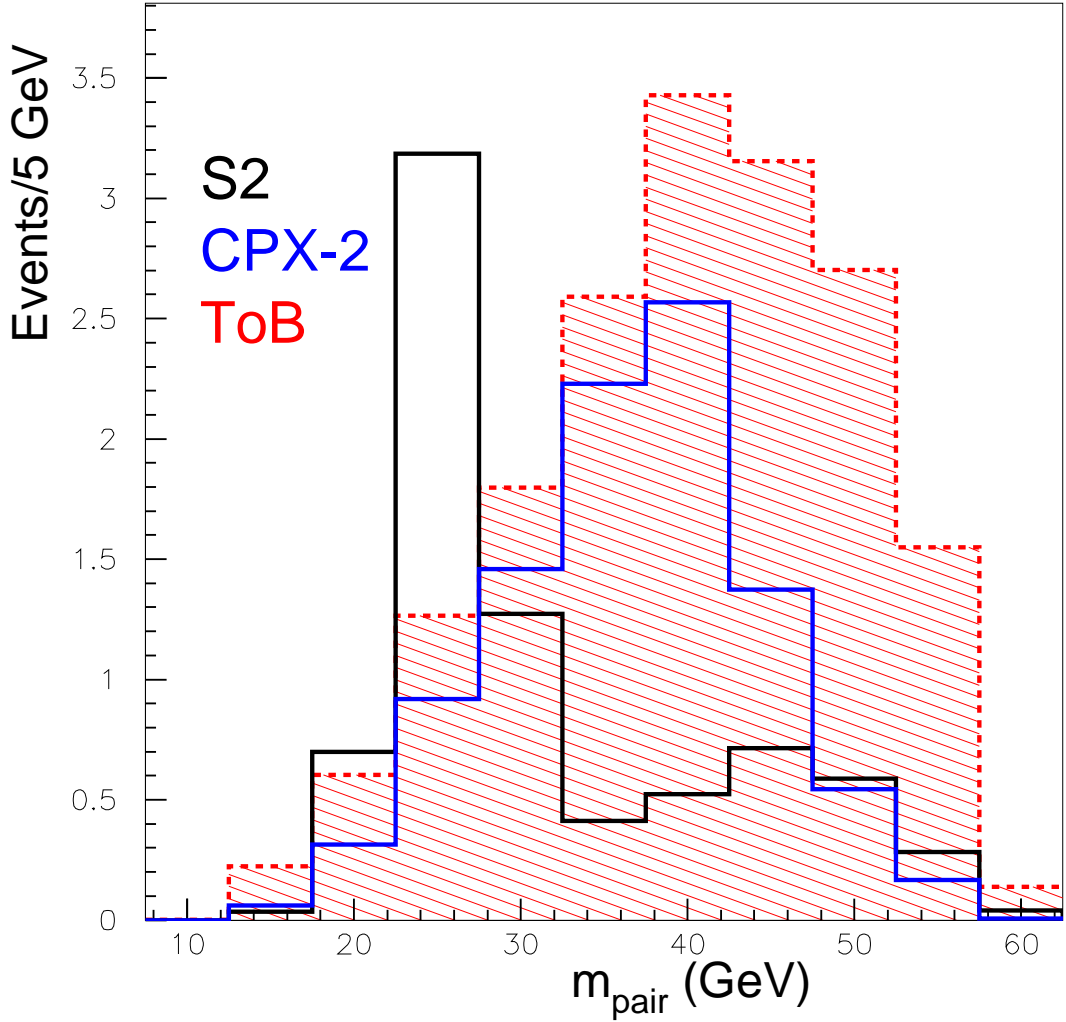}}
\caption{The jet pair invariant mass distribution defined in eq.(\ref{h1}) for
  signal scenarios S2 and CPX-2 (black and blue or dark grey histograms) and
  for the total background (shaded histogram) at the LHC. We have applied the
  acceptance cuts (\ref{lhc-sel}), demanding exactly four reconstructed jets,
  at least three of which are tagged, and required both di--jet invariant
  masses in the optimal pairing to lie between 10 and 60 GeV. The histograms
  give the number of events per bin and per 10 fb$^{-1}$ of integrated
  luminosity.}
\label{lhc_mh1}
\end{center}
\end{figure}

\begin{figure}[h!]
\begin{center}
\rotatebox{0}{\includegraphics[width=14cm]{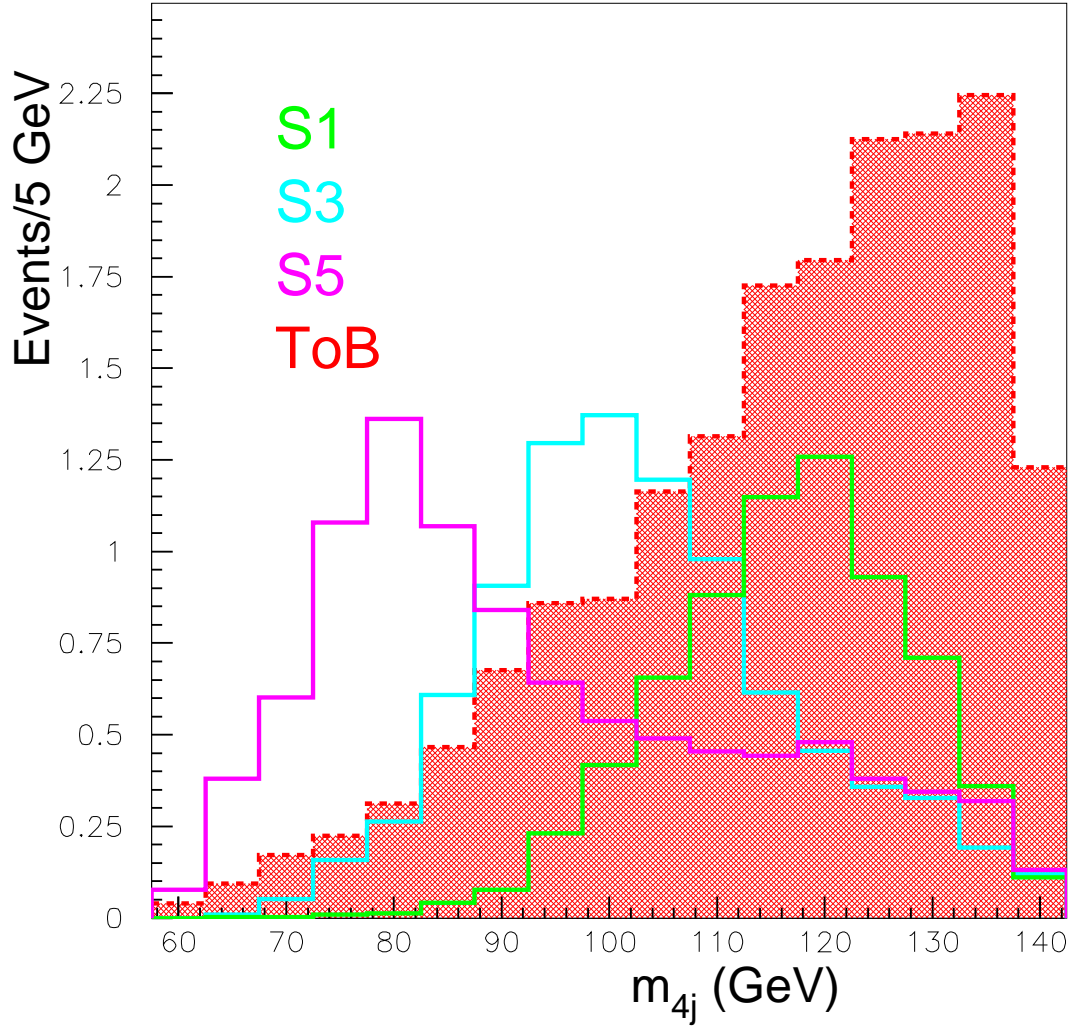}}
\caption{The four--jet invariant mass distribution for signal scenarios S1
  (black), S3 (dark blue or dark grey) and S5 (green or light grey) as well as
  the total background (shaded). We have applied the same cuts of
  Fig.~\ref{lhc_mh1}. The histograms give the number of events per bin and per
  10 fb$^{-1}$ of integrated luminosity. }
\label{lhc_mh2}
\end{center}
\end{figure}

A convincing signal can therefore only be established by detecting
characteristic features in some kinematical distributions. To this end
we consider the $m_{\rm pair}$ and $m_{4j}$ distributions already
discussed for the Tevatron; they are shown in Fig.~\ref{lhc_mh1} and
\ref{lhc_mh2}, respectively. Unfortunately the background shows a peak
in the $m_{\rm pair}$ distribution between 30 and 40 GeV, not far from
the peak of the signal in the scenarios we consider. A tighter cut on
$m_{\rm pair}$ will nevertheless improve the signal--to--background ratio.
Moreover, the four--jet invariant mass distribution of the background
peaks at large values, largely due to the contribution from $t \bar t$
production. At least for scenarios with $h_2$ masses in the lower half
of the ``LEP hole'' region a tighter cut on $m_{4j}$ will therefore
also improve the significance of the signal.

\begin{table}[h!] 
\begin{center}
\begin{tabular}{|c|c|c|c|}
\hline
Scenario  & $S$ & $B$ & ${\cal S}$ \\
\hline
%%%%%%%%%%%%%%%%%%%%%%%%%%%%%%%%%
S1 & 30.3   & 23.7 & 6.22 \\
S2 & 30.4   & 19.8 & 6.83 \\
S3 & 29.8   & 16.1 & 7.43 \\
S4 & 30.1   & 12.7 & 8.45 \\
S5 & 29.4   & 9.70 & 9.44 \\
CPX-1& 35.3 & 21.2 & 7.66 \\
CPX-2& 39.1 & 29.9 & 7.15 \\
%%%%%%%%%%%%%%%%%%%%%%%%%%%%%%%%%
\hline
\end{tabular}
\caption{Final number of signal events $S$, background events $B$ and
  significance ${\cal S}$ at the LHC, defined as ${\cal S} = S /
  \sqrt{B}$, for 60 fb$^{-1}$ of data after all cuts, including the
  double peak requirement (\ref{doublepeak}).}
\label{tab:LHC_signi}
\end{center}
\end{table}

For our final definition of the significance of the signal we therefore
again count the events in the kinematical region defined by the cuts
(\ref{doublepeak}). The results are summarized in
Table~\ref{tab:LHC_signi}, where we assume an integrated luminosity of
60 fb$^{-1}$, corresponding to three years of nominal ``low
luminosity'' running with two experiments. 

As expected from the last two figures, the significance decreases with
increasing $m_{h_2}$. This is almost entirely due to the increase of the
background with increasing $4j$ invariant mass shown in
Fig.~\ref{lhc_mh2}. The number of signal events after all cuts is almost
independent of $m_{h_2}$. We saw in Table~\ref{tab:eventLHC} that scenarios
with smaller $m_{h_2}$ have larger total cross sections. This is only
partially compensated by the increased $b-$tagging efficiency, so that the
signal cross section after the cuts considered in Table~\ref{tab:eventLHC}
still decreases with increasing $m_{h_2}$. However, Fig.~\ref{lhc_mh2} shows
that scenarios with smaller $m_{h_2}$ also have a longer tail of the $4j$
invariant mass distribution towards large values (but below the upper limit of
140 GeV). This results in a reduced efficiency for the final ``double peak''
cut, which happens to almost exactly compensate the $m_{h_2}$ dependence of
the cross section before this cut.

We also see that for given $m_{h_2}$, increasing $m_{h_1}$ reduces the
significance. Recall from table~\ref{tab:eventLHC} that the signal cross
section does not depend much on $m_{h_1}$ after the cuts considered
there. On the other hand, Fig.~\ref{lhc_mh1} again shows poorer
efficiency for passing the final ``double peak'' cut with reduced Higgs
mass, since more of the tail towards larger values (now of the
averaged di--jet invariant mass) is cut away. However, this Figure
also shows an even more rapid increase of the background with
increasing $m_{\rm pair}$.

More importantly, Table~\ref{tab:LHC_signi} shows that after the
double peak cut, the signal always exceeds the background, giving a
final statistical significance of at least 5 standard deviations, and
a signal sample of some 30 events.

This result is somewhat at odds with the previous most detailed
analysis \cite{Han}, which used cuts similar to our's, but did not
include showering, hadronization and the underlying event. The absence
of showering eliminates all $t \bar t$ backgrounds after requiring
$m_{4j} < m_t$. On the other hand, ref.\cite{Han} finds a much larger
$W+4b$ cross section (our process p2), and also finds a sizable
$W+3b+j$ cross section (our process p3). The latter indicates that
these background estimates include processes with $b$ quarks in the
{\em initial} state; otherwise the cross sections for processes with
an odd number of $b$ quarks in the final state would be suppressed by
the square of a small CKM matrix element, making them essentially
negligible (as in our study). In our treatment these reactions are
included in the $W+4b$ background.\footnote{For example, consider the
  process $b u \rightarrow W^+ d b b \bar b$. Treating all $b$
  production explicitly this is described by $g u \rightarrow W^+ d b
  \bar b b \bar b$; this is included in the backward evolution of the
  initial state shower of $\bar d u \rightarrow W^+ b \bar b b \bar
  b$, where the $\bar d$ is created from a $g \rightarrow d \bar d$
  splitting, giving another $d$ quark in the final state.} However,
this does not explain the very large $W+4b$ cross section found in
ref.\cite{Han}, 25 fb after cuts, nearly an order of magnitude larger
than the 3 fb (see Table~\ref{tab:eventLHC}). Moreover, ref.\cite{Han}
finds that the $W+4b$ cross section at the LHC after all cuts is more
than thousand times larger than at the Tevatron. We find this ratio to
be close to 40; this seems more reasonable to us, given that the total
$W+4b$ cross section only increases by about a factor of eleven (as
does the inclusive $W$ production cross section) when going from the
Tevatron to the LHC. On the other hand, process p4, which dominates
our $W+4j$ background, has apparently not been included in
ref.\cite{Han}. Nevertheless our final $S/B$ ratio is more than two
times higher than that of ref.\cite{Han}. However, since the absence
of showering also increases the signal acceptance, the final
significance quoted in ref.\cite{Han} is actually somewhat higher than
our estimate.

\section{Conclusions} 

We analyzed the possibility of observing neutral Higgs bosons at
currently operating hadron colliders in the framework of the CP
violating MSSM. We explored the $\ell jjjj \met$ channel with double,
triple and quadruple $b$ tag, focusing on the region of parameter
space not excluded by LEP searches. We have explicitly considered a
large number of SM backgrounds, breaking up the generic $W+4j$ QCD
backgrounds into many classes, e.g. depending on the number of $b$
quarks in the final state, and carefully treating the production of
additional $b \bar b$ and $c \bar c$ pairs in QCD and $t \bar t$
events. We employed a full hadron--level Monte Carlo simulation using
the {\tt PYTHIA} event generator and its {\tt PYCELL} toy calorimeter. We
carefully implemented $b-$tagging, including mistagging of $c-$jets
or light flavor or gluon jets. 

We first applied this to the Tevatron collider. We found that if we
require three tagged jets, we can only expect about one signal event
per 10 fb$^{-1}$ of integrated luminosity, on a background of about
0.3 events. On the other hand, if we require only double $b-$tag, the
signal increases by a factor of about 4, but the background increases
by two orders of magnitude, again making the signal unobservable.

Going from the Tevatron to the LHC increases the raw signal cross section by
about a factor of 10, whereas some of the important raw background cross
sections increase by two orders of magnitude. We therefore have to demand at
least three $b-$tags. In contrast to previous analyses \cite{Koreans, Han}, we
find $t \bar t$ production to be the biggest background. This is partly due to
the effect of showering. Moreover, we include backgrounds not (explicitly)
considered before, in particular $W b \bar b c j$ final states (where $j$
stands for a light quark) which we find to be the dominant non$-t \bar t$
background at the LHC. Nevertheless, by focusing on events with exactly four
jets, and cutting simultaneously on the average di--jet invariant mass and the
four--jet invariant mass, we found a signal rate above the background, and a
signal significance exceeding 5 standard deviations for an integrated
luminosity of 60 fb$^{-1}$. This luminosity could be accumulated at the end of
``low luminosity'' running of the LHC after summing over both experiments.

Although we improved on earlier analyses in a number of ways, our
treatment of $b-$tagging is not fully realistic. We assumed constant
(mis)tagging probability for jets within a certain rapidity window and
with $E_T$ above 15 GeV, and vanishing probability for all other
jets. Moreover, we assumed that these probabilities factorize,
i.e. can be applied to each jet independent of the rest of the
event. More sophisticated tagging algorithms can directly classify the
entire event as containing a given (minimal) number of $b-$jets.
However, we checked that our simple algorithm reproduces published
results for $t \bar t$ events at the Tevatron; recall that this is one
of our main backgrounds.

Experiments have the possibility to change tagging criteria. This
allows to increase the tagging efficiency at the cost of also
increasing the mistagging probability. Even if (mis)tagging
probabilities indeed factorize, it is not clear that the same
parameter choice should be made for all tags. For example, at the
Tevatron one might try to combine two strong $b-$tags, similar to the
ones we employed, with a weaker one; both the signal rate and the
signal to background ratio should then lie between our results for
double and triple $b-$tag. Recall, however, that the total number of
signal events even with only double $b-$tag is quite small at the
Tevatron. We therefore do not think that such more sophisticated
tagging algorithms can change our pessimistic conclusion regarding the
Tevatron. 

However, at the LHC one could increase the signal to background ratio even
more by requiring a fourth $b-$tag with softer tagging criteria, possibly
simultaneously relaxing the requirement on the number of jets in the event to
increase the statistics. This could be used to confirm the existence of a
signal. On the other hand, using milder criteria already for the third tagged
$b$ is probably not very useful, since the dominant backgrounds contain a $c$
quark which could quite easily be mistagged if too mild tagging criteria are
used. The signal can also be corrobated using $Zh_2$ production with $Z
\rightarrow \ell^+ \ell^-$ \cite{Koreans,Han}. This channel does not receive
significant background from $t \bar t$ production, so the signal to background
ratio should be about two times higher than for the signal we
considered. Unfortunately it also has about five times smaller signal rate,
leaving only about 2 events per 10 fb$^{-1}$ of luminosity at the LHC.

Another concern is the reliability of the background estimates, which are
based on leading order QCD calculations. Higher order corrections to the $t
\bar t$ cross section are known. We did not include them, since our signal
calculation also does not include NLO corrections; moreover, NLO corrections
to $t \bar t$ production are not very large. Since the $W+4j$ cross section is
${\cal O}(\alpha_S^4)$, the leading order estimate suffers from even larger
scale uncertainties than the $t \bar t$ cross section. An almost complete NLO
calculation to $W^- + 4j$ production became available very recently
\cite{berger}. They find moderate {\em negative} NLO corrections.  However,
they use a somewhat smaller renormalization and factorization scale; using
this scale would e.g. increase our $W+4b$ background by about a factor of
1.8. Moreover, their numerical results are for $\sqrt{s} = 7$ TeV, use
significantly stronger cuts on the jet transverse momenta and, most
importantly, do not distinguish the flavor of the jets; it is not at all clear
whether this result carries over to final states containing two, three of four
heavy quarks.\footnote{This is why we do not include known QCD
  \cite{Han:1991ia} and electroweak \cite{Ciccolini:2003jy} corrections to the
  signal cross section.} Recall also that our background requires good control
of the tail of the $t \bar t$ four--jet invariant mass distribution. A careful
validation of background Monte Carlo generators using real data will therefore
be essential before a signal can be claimed.

We conclude that searches for $W h_2$ production with $W \rightarrow
\ell \nu$ and $h_2 \rightarrow h_1 h_1 \rightarrow b \bar b b \bar b$
should be able to close that part of the ``LEP hole'' in parameter
space where $h_1 \rightarrow b \bar b$ decays dominate. The same
search would also probe parts of the parameter space of many
extensions of the MSSM where a heavier Higgs boson can decay into two
lighter bosons, each of which in turn decays into a $b \bar b$
pair. The search will be challenging, but the prize for a successful
search would be well worth the effort: the discovery of not one, but
two Higgs bosons at once!

\subsubsection*{Acknowledgments} 
  
We thank A. Datta, S. Fleischmann, R. Frederix, S. Gonzalez, M. Maity,
F. Maltoni and J. Schumacher for useful discussion. This work was partially
supported by the Bundesministerium f\"ur Bildung und Forschung (BMBF) under
Contract No. 05HT6PDA, by the EC contract UNILHC PITN-GA-2009-237920, and by
the Spanish grants FPA2008-00319, CSD2009-00064 (MICINN) and PROMETEO/2009/091
(Generalitat Valenciana).

\end{document}